\documentclass[aps, prl, superscriptaddress, twocolumn, nofootinbib]{revtex4-2}

\usepackage[T1]{fontenc}
\usepackage{graphicx, nicefrac}
\usepackage{amssymb}
\usepackage{amsmath}
\usepackage[breaklinks=true,colorlinks=true,linkcolor=blue,urlcolor=blue,citecolor=blue]{hyperref}
\usepackage[normalem]{ulem}
\usepackage{color}
\usepackage{subfigure}
\usepackage{physics}
\usepackage{bm}
\usepackage{enumerate}
\usepackage[utf8]{inputenc}
\usepackage{txfonts}
\usepackage[english]{babel}
\usepackage{blindtext,tikz}
\usepackage{siunitx} %not just a theory paper!
\usetikzlibrary{calc}
\usepackage{hhline}

\newcommand{\sigmai}{\sigma_{I}}
\newcommand{\sigmax}{\sigma_{x}}
\newcommand{\sigmay}{\sigma_{y}}
\newcommand{\sigmaz}{\sigma_{z}}

\newcommand{\gate}[1]{#1}

\usepackage{hhline}
\usepackage{amsfonts,mathtools}
\usepackage{placeins}

\makeatletter \renewcommand\d[1]{\ensuremath{%
  \;\mathrm{d}#1\@ifnextchar\d{\!}{}}}
\makeatother

\usepackage{microtype}

\usepackage{siunitx}
\usepackage{chemformula}
\usepackage{braket}

\begin{document}

\title{Characterizing mid-circuit measurements on a superconducting qubit using gate set tomography}

\author{Kenneth~Rudinger}
\email{kmrudin@sandia.gov}
\affiliation{Quantum Performance Laboratory, Sandia National Laboratories, Albuquerque, NM 87185, USA and Livermore, CA 94550, USA}
\author{Guilhem~J.~Ribeill}
\email{guilhem.ribeill@raytheon.com}
\author{Luke~C.~G.~Govia}
\author{Matthew~Ware}
\affiliation{Quantum Engineering and Computing, Raytheon BBN Technologies, 10 Moulton St., Cambridge, MA 02138, USA}
\author{Erik~Nielsen}
\author{Kevin~Young}
\affiliation{Quantum Performance Laboratory, Sandia National Laboratories, Albuquerque, NM 87185, USA and Livermore, CA 94550, USA}
\author{Thomas~A.~Ohki}
\affiliation{Quantum Engineering and Computing, Raytheon BBN Technologies, 10 Moulton St., Cambridge, MA 02138, USA}
\author{Robin~Blume-Kohout}
\author{Timothy~Proctor}
\affiliation{Quantum Performance Laboratory, Sandia National Laboratories, Albuquerque, NM 87185, USA and Livermore, CA 94550, USA}

 \date{\today}
\begin{abstract}
Measurements that occur within the internal layers of a quantum circuit --- \emph{mid-circuit measurements} --- are an important quantum computing primitive, most notably for quantum error correction. Mid-circuit measurements have both classical and quantum outputs, so they can be subject to error modes that do not exist for measurements that terminate quantum circuits. Here we show how to characterize mid-circuit measurements, modelled by quantum instruments, using a technique that we call \emph{quantum instrument linear gate set tomography} (QILGST). We then apply this technique to characterize a dispersive measurement on a superconducting transmon qubit within a multiqubit system. By varying the delay time between the measurement pulse and subsequent gates, we explore the impact of residual cavity photon population on measurement error. QILGST can resolve different error modes and quantify the total error from a measurement; in our experiment, for delay times above $\SI{1000}{ns}$ we measured a total error rate (i.e., half diamond distance) of $\epsilon_{\diamond} = 8.1 \pm 1.4 \%$, a readout fidelity of $97.0 \pm  0.3\%$, and output quantum state fidelities of $96.7 \pm 0.6\%$ and $93.7 \pm 0.7\%$ when measuring $0$ and $1$, respectively.
\end{abstract}

\maketitle
\begin{figure}[b!]
\includegraphics[width=8.7cm]{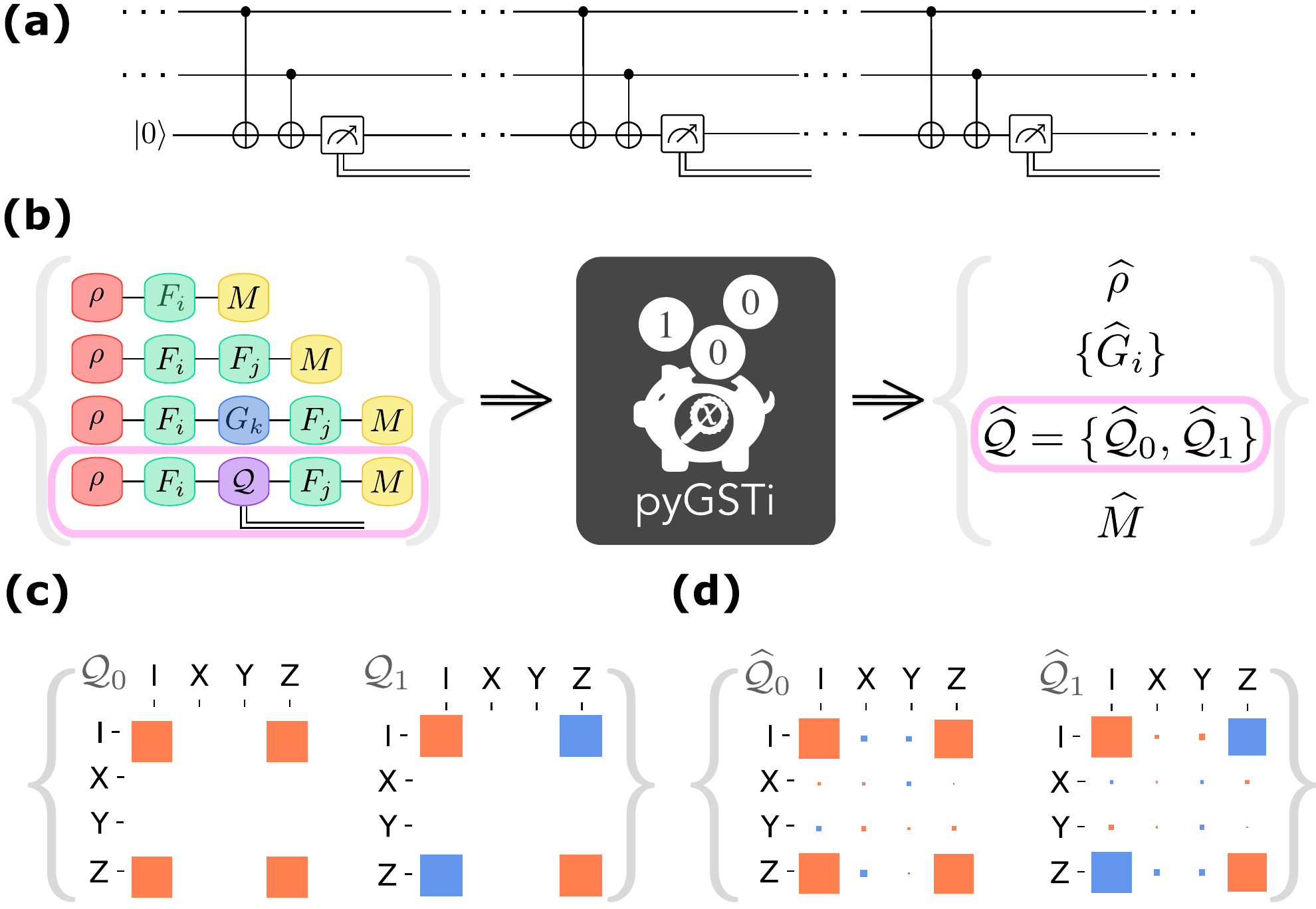}
\caption{\textbf{Characterizing mid-circuit measurements.} Many quantum computing primitives require mid-circuit measurements, as illustrated by \textbf{(a)} a repeated parity-check circuit. \textbf{(b)} Our QILGST protocol for characterizing a mid-circuit measurement, as part of a complete gate set ($\mathcal{G}=\{\rho, G_i, Q, M\}$). The mid-circuit measurement is modelled by a quantum instrument $Q=\{Q_i\}$, which consists of a process matrix for each measurement outcome. QILGST consists of (1) running circuits that enable process tomography on $Q$, alongside the circuits of standard GST \cite{Greenbaum2015-tr, Blume-Kohout2017-no, Nielsen2020-lt}; and (2) closed-form matrix inversion or maximum likelihood estimation (as implemented in \texttt{pyGSTi} \cite{Nielsen2020-lu}) to obtain a self-consistent reconstruction of the gate set ($\widehat{\mathcal{G}}=\{\widehat{\rho}, \widehat{G}_i, \widehat{Q}, \widehat{M}\}$). Additions to standard GST are circled in pink. We applied QILGST to characterize a dispersive $\sigmaz$ basis measurement on a transmon qubit. The $\textbf{(c)}$ target and $\textbf{(d)}$ estimated QI from our experiment, which has a readout fidelity of $97.0\% \pm 0.3\%$ and a total error rate of $\epsilon_{\diamond} = 8.1 \pm 1.4 \%$. Each orange (blue) square represents a positive (negative) real number whose magnitude is proportional to the square’s area.}
\label{fig:schematic}
\end{figure}

Gate-model quantum computers perform computations by executing sequences of quantum operations, known as quantum circuits. Quantum computations can be performed with circuits that contain only qubit initialization, reversible logic gates, and \emph{terminating measurements} \cite{DiVincenzo2000-ei} --- meaning measurements that occur at the circuit's end, and convert the quantum information stored in the qubits into classical bits. However, circuits can also contain \emph{mid-circuit measurements} that extract information from the qubits and alter their state, but do not destroy the qubits nor necessarily collapse their state entirely. High-fidelity mid-circuit measurements are essential for quantum error correction (QEC) \cite{Shor1995,knill2005quantum, divincenzo2007effective, landahl2011fault, Fowler2012-ea, devitt2013quantum} --- a \emph{parity check} or \emph{stabilizer} measurement (Fig.~\ref{fig:schematic}a) must extract information about a specific multiqubit observable, while not disturbing the quantum information stored in the logical subspace --- and they also have applications to error mitigation and implementation of near-term algorithms \cite{Riste2020, Corcoles2021-ht, Urbanek2020-nm, Holmes2020-wl, Urbanek2020-nm, paetznick2013repeat}. Mid-circuit measurements, however, admit failure modes that do not exist for terminating measurements. Techniques for precise, reliable \emph{characterization} of mid-circuit measurements are therefore urgently needed. In this Letter, we introduce a protocol (Fig.~\ref{fig:schematic}b) for comprehensive and self-consistent characterization of a full set of logic operations that includes mid-circuit measurements --- which we call \emph{quantum instrument linear gate set tomography} (QILGST). We then use QILGST to study single-qubit dispersive measurements on a superconducting transmon processor (Fig.~\ref{fig:schematic}d).

Techniques for assessing the performance of quantum logic operations can be divided into benchmarking and characterization. Benchmarks quantify the overall performance of operations \emph{in situ} on representative tasks, and mid-circuit measurements can be (and have been) benchmarked using QEC (and components thereof) \cite{Chen2021-nv, andersen2020repeated, erhard2021entangling, negnevitsky2018repeated, rosenblum2018fault, elder2020high, bultink2020protecting, fowler2014scalable, takita2016demonstration, willsch2018testing, Combes2017-ns} or algorithm \cite{Corcoles2021-ht} circuits. However, identifying specific error modes, predicting their impact, and mitigating or eliminating them requires detailed characterization.  This is commonly done by tomography, which means estimating a model for the operation. Terminating measurements are modeled by \emph{positive operator-valued measures} (POVMs) and can be estimated by \emph{quantum detector tomography} \cite{Luis1999-lw, Lundeen2009-jo, Fiurasek2001-yl, DAriano2004-rb, Izumi2020-db}, but only if precalibrated input states and gates are available.  Gate set tomography (GST) \cite{merkel2013self, Greenbaum2015-tr, Blume-Kohout2017-no, Nielsen2020-lt} removes this requirement, enabling estimation of POVMs self-consistently together with initialization and logic gates.  We now show how to extend GST to gate sets that also include mid-circuit measurements, represented as \emph{quantum instruments} \cite{Davies1970-js}. Prior works \cite{Miklin2020-pn, Mohan2019-xe, Wagner2020-hl, Blumoff2016-gy} show how to perform self-testing or tomography of quantum instruments, but, to the best of our knowledge, this is the first protocol for complete \emph{and} self-consistent tomography of mid-circuit measurements.

\vspace{0.2cm}
\noindent
\textbf{Quantum instruments ---} Quantum instruments (QIs) \cite{Davies1970-js} are the natural mathematical model of mid-circuit measurements for tomography. In tomography, a quantum processor's state is represented by a $d\times d$ density matrix, where $d$ is the (intended) dimension of the processor's Hilbert space. Gates are represented by superoperators that act linearly on density matrices and terminating measurements by POVMs that map density matrices to probability distributions. All of these objects are completely positive (CP) and trace preserving (TP) \emph{quantum processes}. They differ only by their input and output spaces. States (density matrices) describe initialization; they map a trivial space \emph{into} the $d^2$-dimensional space of mixed states. Gate superoperators map that space to itself. POVMs map quantum states to distributions over outcomes. QIs are simply another special instance: they are processes with a quantum input, and quantum \emph{and} classical outputs.  This describes a mid-circuit measurement, combining the features (and outputs) of a POVM and a gate.  The simplest representation of an $m$-outcome QI $\gate{Q}$ is as a set of $m$ CP maps  $\gate{Q} = \{\gate{Q}_0,\ldots,\gate{Q}_{m-1}\}$ whose sum $\sum_iQ_i$ is a TP map. The QI maps $\rho$ to a joint quantum-classical state $\{(p_i,\rho_i)\}_{i=0}^{m-1}$, where $p_i = \text{Tr}(\gate{Q}_i[\rho])$ is the probability of observing outcome $i$ and $\rho_i=\gate{Q}_i[\rho]/p_i$ is the output state \emph{conditional on} observing $i$. As with gates, each $Q_i$ can be represented using a $d^2 \times d^2$ process matrix (see Fig.~\ref{fig:schematic}c for an example, with matrix elements defined by $[Q_j]_{kl} = \text{Tr}(\sigma_k Q_j[\sigma_l])$ for $k,l=I,x,y,z$).

Quantum instruments can model errors in mid-circuit measurements that POVMs cannot.  A POVM cannot represent a mid-circuit measurement at all, because POVMs have strictly classical outputs. Any POVM, however, can be ``promoted'' to a limited kind of QI called a ``measure-and-prepare'' process \cite{Horodecki2003-fm}, by following it with a conditional re-initialization (i.e., upon observing $i$, $\rho_i$ is prepared). A measurement that is describable as a measure-and-prepare process can be characterized with existing methods (e.g., GST), but measure-and-prepare processes cannot describe all mid-circuit measurements. Measure-and-prepare processes destroy all entanglement with other quantum systems \cite{Horodecki2003-fm}, but, e.g., QEC parity-checks should preserve specific kinds of inter-qubit entanglement perfectly. Conversely, mid-circuit measurements \emph{designed} to be measure-and-prepare processes can easily fail in ways that cannot be modeled without a general QI \cite{supplement}. QIs can model and describe all Markovian errors in mid-circuit measurements, and our goal is to reconstruct (from data) the QI that describes an experimental mid-circuit measurement.

\begin{figure}[t!]
\includegraphics[width=8cm]{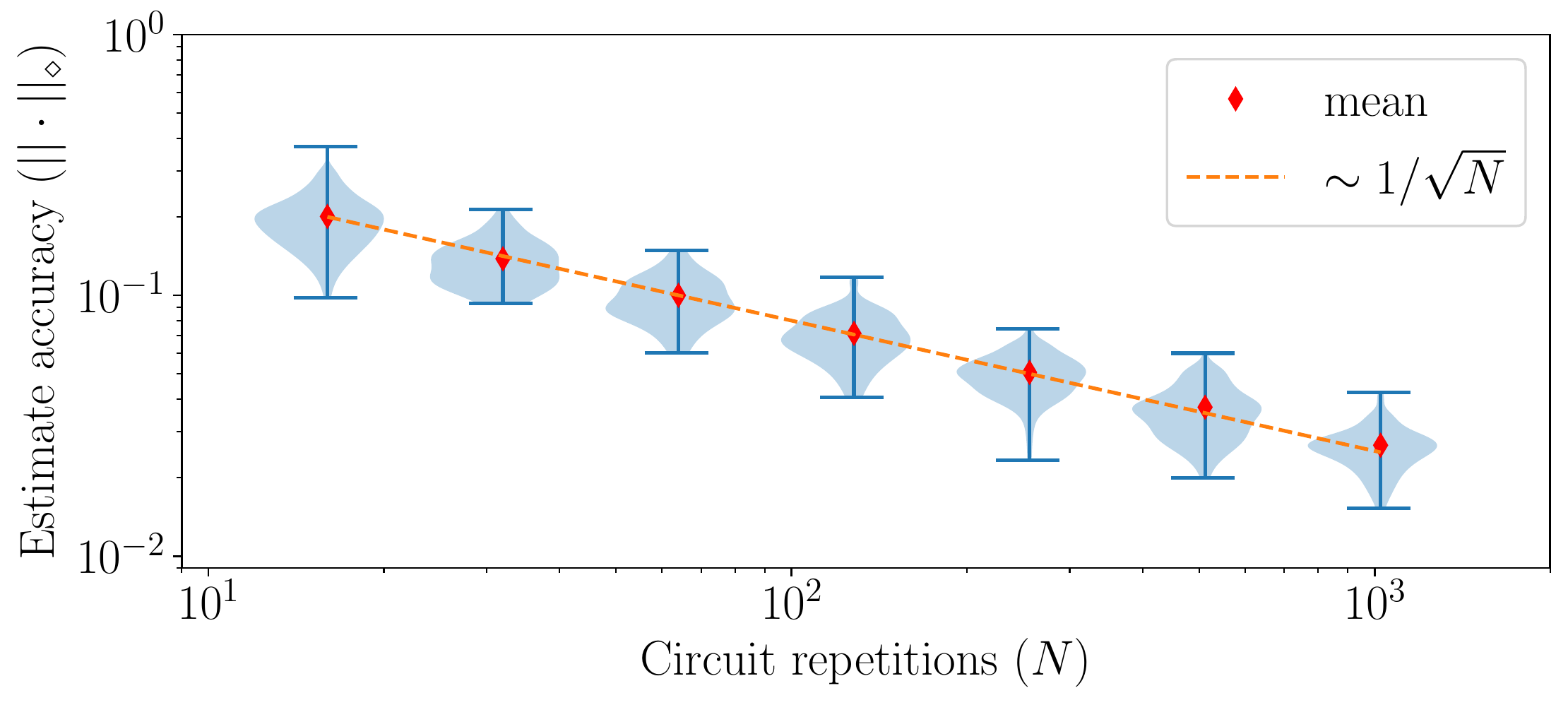}
\caption{\label{fig:simulations}\textbf{QILGST accurately characterizes mid-circuit measurements.} We simulated single-qubit QILGST under a variety of error models \cite{supplement} and computed the accuracy of the estimated QI $\widehat{Q}$. This plot shows the estimation accuracy, measured by half the diamond distance ($\epsilon_{\diamond})$ \cite{Aharonov1998-ov} between $\widehat{Q}$ and the QI used to generate the data, versus the number of samples drawn from each circuit ($N$). Each point (violin plot) is the mean (distribution) of the estimation inaccuracy from simulating QILGST under 100 different error models. The accuracy scales as $O(\nicefrac{1}{\sqrt{N}})$, which is the expected shot noise scaling.}
\end{figure}

\vspace{0.2cm}
\noindent
\textbf{GST with quantum instruments ---} GST \cite{Nielsen2020-lt, Blume-Kohout2017-no, Greenbaum2015-tr} simultaneously and self-consistently reconstructs all the elements of a gate set $\mathcal{G}$ --- containing an initialization $\rho$, two or more logic gates $\{\gate{G}_i\}$, and a terminating measurement $M$. It specifies (1) an experiment design (a set of circuits to be performed) and (2) analysis procedures for transforming data into an estimate of the gate set.  Several variants exist \cite{Nielsen2020-lt}; here, we adapt \emph{linear-inversion} GST (LGST) to gate sets containing mid-circuit measurements. LGST is similar to process tomography \cite{OBrien2004-tr, Poyatos1997-mz, Chuang1997-vf}, with three key innovations: (1) to tomograph each gate $\gate{G}_i$, the experiment includes all circuits of the form $F^p_j \gate{G}_i F^m_k$ where the \emph{fiducial circuits} $\{F^p_j\}_{j=1}^{n_p}$ and $\{F^m_k\}_{k=1}^{n_m}$ produce informationally complete ensembles of states and terminating measurements, respectively, using only gates in $\mathcal{G}$; (2) the experiment includes circuits for process tomography on the null operation (the $F^p_jF^m_k$ circuits); and (3) systematic errors are removed using the inverse of the tomographed null operation \cite{Greenbaum2015-tr, Nielsen2020-lt}. To extend LGST to a gate set containing a mid-circuit measurement, represented by a QI $\gate{Q}$ \footnote{We only consider a single QI here; extending to multiple QIs in the same gate set is trivial.}, we simply add all circuits of the form $F^p_j \gate{Q} F^m_k$ to the LGST experiment (Fig. \ref{fig:schematic}b); these circuits output a result from both the mid-circuit \emph{and} terminating measurement.

Analyzing QILGST data presents one complication.  Whereas each gate $\gate{G}_i$ is represented by a single CPTP map, a QI defines a \emph{set} of CP maps $\{\gate{Q}_0,\cdots, \gate{Q}_{m-1}\}$. Which $\gate{Q}_i$ appears in any given run of the circuit is not controllable; it's determined by the mid-circuit measurement's outcome. To reconcile this with the LGST analysis, we represent the QI by a $md^2 \times d^2$ process matrix
\begin{equation}
    \gate{Q} = \left(\begin{array}{c}\gate{Q}_0, \ldots,  \gate{Q}_{m-1}\end{array}\right)^\intercal,
\end{equation}
which is a CPTP map. The $m$ blocks correspond to copies of the quantum state space, indexed by the measurement's classical outcome.  So whereas the LGST linear inversion algorithm for a gate $\gate{G}$ begins with a $d^2\times d^2$ matrix of directly measured probabilities $\tilde{\gate{G}}_{kj}$ \cite{Nielsen2020-lt} --- where the row $k$ labels a final measurement setting and the column $j$ labels a preparation setting --- the corresponding algorithm for a QI $\gate{Q}$ starts with an $md^2\times d^2$ matrix of probabilities $\tilde{\gate{Q}}_{kj}$ where $k$ labels a final measurement setting \emph{and} which outcome of $\gate{Q}$ was observed.  With this modification, the LGST algorithm can be directly applied, with the matrix elements of $\gate{Q}$ estimated to the same absolute precision as those of a gate $\gate{G}$.

We call this protocol \emph{quantum instrument linear GST} (QILGST). It requires only about 100 circuits to characterize a single-qubit gate set including 2-3 gates and a QI $\gate{Q}$.  For all analyses in this Letter, we used numerical maximum likelihood estimation (implemented in \texttt{pyGSTi} \cite{Nielsen2020-lu,Nielsen2019-wi}), instead of closed-form linear inversion. This yields higher accuracy by accounting for heteroskedasticity in the data \footnote{It also allows us to constrain the gates to be CPTP. We constrain the QI to be TP but not CP.}. Data analysis for single-qubit QILGST takes a few seconds on a modern laptop. To verify the correctness of QILGST, we simulated it for a variety of error models \cite{supplement}; QILGST correctly reconstructs the QIs (Fig.~\ref{fig:simulations}).

\vspace{0.2cm}
\noindent
\textbf{Quantifying errors in a mid-circuit measurement ---}
Running QILGST yields estimates of all the gates \emph{and} an estimated QI for the mid-circuit measurement.  Like all GST estimates, it has a gauge freedom \cite{Nielsen2020-lt}, which we fix by numerically varying the gauge to minimize the discrepancy between the estimated gates and their targets (\emph{gauge optimization} \cite{Nielsen2020-lt}).  We denote the gauge-optimized estimate of $\gate{Q}$ by $\widehat{\gate{Q}}$.  The estimated $\widehat{\gate{Q}}$ can be compared to the ideal ``target'' QI ($\gate{Q}_{\mathrm{target}}$), to quantify the errors in the mid-circuit measurement. As with gates, a mid-circuit measurement can display a variety of distinct errors, e.g., measuring the wrong observable, scrambling the classical information in the measurement result, or creating the wrong post-measurement quantum state. So it is common to summarize the quality of a logic operation with a metric such as fidelity or diamond norm.  Fidelity between QIs \cite{Magesan2011-hc} is difficult to interpret because of the joint quantum/classical output, but the diamond distance error $\epsilon_\diamond = \frac12\|\widehat{\gate{Q}}-\gate{Q}_{\mathrm{target}}\|_{\diamond}$ \cite{Aharonov1998-ov} is well-defined and is a tight upper bound on the change in any experimental probability induced by errors in $\gate{Q}$. %

\begin{figure}[t!]
\includegraphics[width=8.6cm]{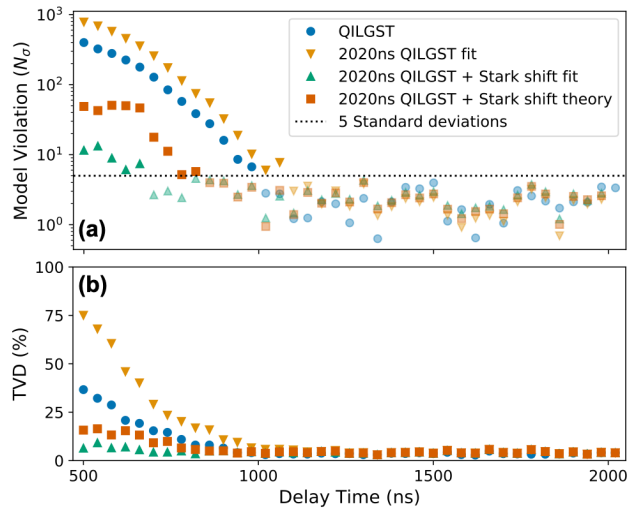}
\caption{\textbf{Characterizing non-Markovian errors in a mid-circuit measurement on a superconducting qubit.} The evidence for unmodelled error as a function of the delay time ($t_d$) between the measurement pulse and subsequent operations, for four different models estimated from the data using QILGST. The evidence for unmodelled error is quantified by \textbf{(a)} the number of standard deviations ($N_{\sigma}$) of model violation, and \textbf{(b)} the largest total variation distance (TVD) between the model's prediction and the data, for the 36 circuits containing a mid-circuit measurement. The gate set estimated by QILGST (circles) does not accurately describe the data for short delay times, indicating non-Markovian errors. This additional error can be modelled by combining QILGST's model estimated from the $\SI{2020}{ns}$ data --- which, alone, is not consistent with the small $t_d$ data (down-triangles) --- with a decaying Stark shift error on the gates that follow the measurement (squares and up-triangles).}
\label{fig:results}
\end{figure}

\vspace{0.2cm}
\noindent
\textbf{Experiments ---} We used QILGST to study mid-circuit measurements on a transmon qubit within a five-qubit device. We performed dispersive measurements through a microwave cavity coupled to the qubit using standard circuit QED~\cite{Blais:2004} methods. We achieved a high readout fidelity of $\sim\SI{96}{\percent}$ using a JTWPA amplifier \cite{OBrian:2014}, with a \SI{1}{\micro\second} long measurement pulse resonant with the qubit ground-state shifted cavity frequency, that is subsequently digitized and integrated using a matched-filter kernel \cite{ryan2015tomography}. The measurement pulse amplitude, measured through the qubit Stark shift \cite{McClure2016}, created an average cavity population of $\bar{n} = \num{122}$ for the qubit ground state (and substantially less for the excited state), well below the critical photon number $n_c = \alpha\Delta/[4\chi(\alpha+\Delta)] = \num{340}$.  Further device and experimental details can be found in the Supplemental Material \cite{supplement}.

The gate set $\mathcal{G}$ consisted of $\nicefrac{\pi}{2}$ rotations around the $\sigmax$ and $\sigmay$ axes, an idle operation, mid-circuit and terminating measurements in the $\sigmaz$ basis, and state preparation in $\ket{0}$ (implemented by a \SI{500}{\micro\second} idle reset). The mid-circuit measurement's target QI (Fig.~\ref{fig:schematic}c) is $\gate{Q}_{\text{target}} =\{\gate{Q}_{\text{target},0},\gate{Q}_{\text{target},1}\}$ where
\begin{equation}
\gate{Q}_{\text{target},k}[\rho] =  \text{Tr}\left[\frac{1}{2}\left(\sigmai + (-1)^k\sigmaz\right)\rho\right]\left(\sigmai + (-1)^k\sigmaz\right).
\label{eq:zQI}
\end{equation}
For this gate set, there are 128 QILGST circuits \footnote{The preparation and measurement fiducials were the same. They were $G_{x}$, $G_{y}$, $G_{x}G_{x}$, $G_{x}G_{x}G_{x}$, and $G_{y}G_{y}G_{y}$, where $G_{k}$ denotes a $\sigma_k$ rotation by $\nicefrac{\pi}{2}$.}, 36 of which contain a mid-circuit measurement. We ran the QILGST experiment (with $N=1024$ circuit repetitions) multiple times; for each run of the experiment we used a different time delay $t_{d}$ between the mid-circuit measurement pulse and subsequent operations, with $\SI{500}{\ns} \leq t_d \leq \SI{2020}{\ns}$ \footnote{The \SI{500}{\ns} minimum time between a measurement pulse and subsequent gate operation is a limitation of our custom digitizer firmware.}. This produced a QILGST dataset $D(t_d)$ for each $t_d$. We applied the QILGST analysis to each  $D(t_d)$ independently, producing an estimated gate set $\widehat{\mathcal{G}}(t_d)$ for each $t_d$.

The gate set estimated by QILGST will accurately describe the data if the errors on all the operations are Markovian. Non-Markovian errors are common however \cite{Wan2019-rn, Blume-Kohout2017-no, Blume-Kohout2020-fl, Rudinger2019-ua, Proctor2020-iz, Bylander2011-cf}, so we check whether $\widehat{\mathcal{G}}(t_d)$ is consistent with $D(t_d)$ using the log-likelihood ratio test statistic $\lambda_{\rm LLR}$ \cite{Blume-Kohout2017-no, Rudinger2019-ua, nielsen2021efficient}. This $\lambda_{\rm LLR}$ is $N_{\sigma} \sim 400$ standard deviations above its expected value (under the hypothesis that the QILGST model is true) when $t_d = \SI{500}{ns}$, but $N_{\sigma} \lesssim 5$ if $t_d \geq \SI{1020}{ns}$ (Fig.~\ref{fig:results}a, circles). Therefore, short delay times are causing non-Markovian errors. We quantify the size of the unmodelled effect by the total variation distance (TVD) between the probabilities predicted by $\widehat{\mathcal{G}}(t_d)$ and the observed frequencies $D(t_d)$ \cite{Rudinger2019-ua}. The maximum TVD for the 36 QI-containing circuits is large at the shortest delay times ($37\%$ at $t_d = \SI{500}{ns})$, but is small ($< 6\%$) for $t_d \geq \SI{1020}{ns}$ (Fig.~\ref{fig:results}c, circles). We attribute this large non-Markovian error at short delay times to the residual photon population in the measurement cavity (we measure the relaxation rate of the cavity to be $\kappa^{-1} = \SI{242}{\ns}$), as we did not perform active reset of the cavity state \cite{McClure2016, Bultink2018}.

Before investigating the non-Markovian effects observed for $t_d \leq \SI{900}{ns}$, we present the results of QILGST at long delay times ($t_d \geq \SI{1020}{ns}$), where the estimated gate sets \emph{do} accurately describe the data. When $t_d \geq \SI{1020}{ns} \approx 4.2/\kappa$ the cavity photon population is negligible, so we expect that the only difference in the gate set across those $t_d$ values will be a small increase in relaxation errors, in the mid-circuit measurement's preparation of $\ket{1}\!\bra{1}$, for longer delay times. As $\widehat{\mathcal{G}}(\SI{2020}{ns})$ accurately models the data for all $t_d \geq \SI{1120}{ns}$ (Fig. \ref{fig:results}, down-triangles), we focus on $\widehat{\mathcal{G}} \equiv \widehat{\mathcal{G}}(\SI{2020}{ns})$ and $\widehat{\gate{Q}} \equiv \widehat{\gate{Q}}(\SI{2020}{ns})$, our estimate of the mid-circuit measurement's QI. Figs.~\ref{fig:schematic}d and \ref{fig:schematic}c show the process matrices of $\widehat{\gate{Q}}$ and those of the target $\gate{Q}_{\text{target}}$, respectively. We find that the total error in $\widehat{\gate{Q}}$ is $\epsilon_{\diamond}= 8.1 \pm 1.4\%$ (error bars are at the $2\sigma$ level).  This metric quantifies \emph{all} errors in the measurement, including readout errors \emph{and} errors in the quantum states produced by the measurement.

To verify that the QILGST estimate is consistent with standard techniques for partially characterizing a measurement, we calculate the readout fidelity $F = \frac{1}{2}[P_{0\mid 0} + P_{1 \mid 1}]$ of the mid-circuit measurement, where $P_{0 \mid 0}$ (resp., $P_{1 \mid 1}$) is the (marginal) probability of reading out 0 (resp., 1) in the mid-circuit measurement of the prepare-measure-measure (resp., prepare-$\pi$-pulse-measure-measure) circuit. These two circuits are part of the QILGST experiment, so we can both predict $F$ from $\widehat{\mathcal{G}}$ and compare this to the observed frequencies in $D(\SI{2020}{ns})$. The predicted and directly observed values are $F = 97.0 \pm 0.3 \%$ and $F = 97.3 \pm 0.4\%$, respectively, which are consistent with each other \emph{and} with readout fidelity measurements conducted independently of the QILGST experiment \cite{supplement}.

The readout fidelity does not quantify all of the error in the mid-circuit measurement ($F=97\%$ whereas $\epsilon_{\diamond} = 8\%$). From $\widehat{Q}$'s two process matrices $\{\widehat{Q}_0, \widehat{Q}_1\}$ (Fig.~\ref{fig:schematic}d) we can ascertain the types of errors that are occurring, and quantify their size. As it ideally should, the measurement destroys all coherence between $\ket{0}\!\bra{0}$ and $\ket{1}\!\bra{1}$. This is because, to within statistical uncertainty, $\widehat{Q}_i[\sigmax]=\widehat{Q}_i[\sigmay]=0$ for both $i=0$ and $i=1$ (i.e., only the corner elements of the matrices in Fig.~\ref{fig:schematic}d are inconsistent with zero). $\widehat{Q}$ is therefore entirely described by the probabilities $p_{i \mid j} = \text{Tr}(\widehat{Q}_i[\ket{j}\!\bra{j}])$ and output states $\rho_{i \mid j} = \widehat{Q}_i[\ket{j}\!\bra{j}]/p_{i \mid j}$. We find that $p_{0 \mid 0}=99.7 \pm 0.6\%$ and $p_{1 \mid 1}=99.0 \pm 0.6\%$ (these probabilities imply a readout fidelity of $\tilde{F} = 99.3\pm0.4\%$, which differs from $F$ above --- but it is not inconsistent, as $F$ includes contributions from errors in the state input into the mid-circuit measurement, whereas $\tilde{F}$ does not \footnote{$\tilde{F}$ is also not directly measurable; it is not gauge-invariant.}). We find that $\rho_{i \mid i } = \sigmai + z_i \sigmaz$ where $z_0 = 0.93$ and $z_1 = -0.86$, implying state fidelities between $\rho_{i \mid i }$ and the ideal preparations $\ket{i}\!\bra{i}$ of $96.8 \pm 0.6 \%$ and $93.7  \pm 0.8 \%$, for $i=0$ and $i=1$, respectively. The error in the output quantum state is therefore the dominant error in the mid-circuit measurement. This error is not quantified by readout fidelity, and it cannot be measured by detector tomography. The probability for the excited state to decay during the full $\SI{3.02}{\mu s}$ measurement and delay time is $\sim4.2\%$, so there is an additional source of 2-3\% error in the measurement operation, which we conjecture is due to effects beyond the dispersive model \cite{Boisson2008, Boisson2010, Govia2015, Govia:2016, Khezri2016, Huembeli2017}.

A likely source of the observed non-Markovian error when $t_d < \SI{1020}{ns}$ is residual photons in the cavity, which induce a Stark shift in the qubit frequency.  This causes a $\delta \sigma_z$ Hamiltonian error in all post-measurement gates with (1) $\delta$ decaying over time, and (2) $\delta$ depending on the result of the mid-circuit measurement. In the context of tomography, this is a non-Markovian effect --- it cannot be modeled by a single CPTP map for each gate. To test whether the Stark shift explains the data, we constructed a model $\widehat{G}_{\rm stark}$ that is the same as $\widehat{G}(\SI{2020}{ns})$ except that we added in errors that model the Stark shift. We replaced $\widehat{G}_{k}$ with
\begin{equation}
\widehat{G}_{k}(\alpha, r, i, m) = \exp[\log(\widehat{G}_{k}) + \alpha_i(t_d) \exp(-mr_i) \mathcal{Z}],
\end{equation}
for $k=x,y$, where $\mathcal{Z}[\rho] = -i\sigmaz \rho + i\rho\sigmaz$ is the generator of $\sigmaz$ rotations, $m=0,1,\dots$ indexes the number of gates since the mid-circuit measurement, $i$ is the outcome of the mid-circuit measurement, $r_i$ is the decay rate of the Stark shift, and $\alpha_i(t_d)$ is the initial phase error with the delay time $t_d$. Both $r_i$ and $\alpha_i(t_d)$ can be fully described by dispersive theory and independent device characterizations \cite{supplement}. This model explains the majority of the discrepancy between the the QILGST fits and the data for $t_d < \SI{1100}{ns}$ (Fig.~\ref{fig:results}, squares). With \emph{zero} fit parameters, at $t_d = \SI{500}{ns}$ we have decreased $N_{\sigma}$ by almost an order of magnitude, and the maximum TVD from 80\% to 15\%. This is strong evidence that the main source of the non-Markovianity is this decaying Stark shift. 

This model does not, however, entirely explain the data at the shortest delay times. This could be due to inaccuracies in the device parameter characterization, or effects beyond dispersive theory \cite{Boisson2008, Boisson2010, Govia2015, Govia:2016, Khezri2016, Huembeli2017}. To test the first hypothesis, we \emph{fit} the four parameters $\alpha_0(t_d)$, $\alpha_1(t_d)$, $r_0$, and $r_1$ to the data at each $t_d$. This final model is almost consistent with the data (Fig.~\ref{fig:results}, up-triangles), and its optimized parameter values predict behavior close to that predicted by the independently measured device parameters at long delay times \cite{supplement}.  This demonstrates how QILGST can be combined with device physics to develop and validate microscopic models of device dynamics, while also suggesting that additional physics is needed to fully describe dispersive measurements on superconducting qubits. 

\vspace{0.2cm}
\noindent
\textbf{Discussion ---} Quantum computing experiments that rely on well-calibrated mid-circuit measurements are becoming increasingly preeminent \cite{Chen2021-nv, andersen2020repeated, erhard2021entangling, negnevitsky2018repeated, rosenblum2018fault, elder2020high, bultink2020protecting, fowler2014scalable, takita2016demonstration, willsch2018testing}, and techniques like QILGST  will be essential for characterizing these operations. The most striking features of our experimental results are the non-Markovianity of the mid-circuit measurement at short delay times, and the large error in the post-measurement state even with the longest post-measurement delay. These effects could not have been discovered and quantified using quantum detector tomography, randomized benchmarking, or readout fidelity measurements, and they suggest that active cavity and qubit reset \cite{McClure2016} will be critical for low-error mid-circuit measurements on superconducting qubits. As with standard tomographic methods, the number of circuits required for QILGST scales exponentially with the number of qubits. However, QILGST could potentially be combined with recent advances in many-qubit GST \cite{Nielsen2020-co, govia2020bootstrapping} to obtain polynomial resource scaling. By enabling complete characterizations of, e.g., many-qubit syndrome extraction cycles, this would provide invaluable insight into experimental QEC.

\vspace{0.2cm}
\noindent
\textbf{Acknowledgements ---} The authors would like to thank G. E. Rowlands for assistance with experimental software infrastructure, A. Wagner for assistance with device fabrication,  B. Hassick for experimental assistance, M. da Silva, J. Gamble, and C. Granade for useful discussions, and W. D. Oliver for providing the JTWPA. This material is based upon work supported by the U.S. Department of Energy, Office of Science, Office of Advanced Scientific Computing Research through the Quantum Testbed Program; the Office of the Director of National Intelligence (ODNI), Intelligence Advanced Research Projects Activity (IARPA); and by the U.S. Army Research Office under Contract No: W911NF-14-C-0048. Sandia National Laboratories is a multi-program laboratory managed and operated by National Technology and Engineering Solutions of Sandia, LLC., a wholly owned subsidiary of Honeywell International, Inc., for the U.S. Department of Energy's National Nuclear Security Administration under contract DE-NA-0003525. Fabrication of the devices presented in this paper was
partially conducted at the Harvard Center for Nanoscale Systems, a member of the National Nanotechnology Coordinated Infrastructure Network (NNCI), which is supported by the National Science Foundation under NSF award no. 1541959. All statements of fact, opinions, findings and conclusions or recommendations expressed in this material are those of the authors and do not necessarily reflect the official views or policies of the U.S. Army Research Office, the U.S. Department of Energy, IARPA, the ODNI, or the U.S. Government. This article does not contain technology or technical data controlled under either the U.S. International Traffic in Arms Regulations or the U.S. Export Administration Regulations.

\bibliography{Bibliography_and_supplement-2.bib}

\begin{thebibliography}{76}%
\makeatletter
\providecommand \@ifxundefined [1]{%
 \@ifx{#1\undefined}
}%
\providecommand \@ifnum [1]{%
 \ifnum #1\expandafter \@firstoftwo
 \else \expandafter \@secondoftwo
 \fi
}%
\providecommand \@ifx [1]{%
 \ifx #1\expandafter \@firstoftwo
 \else \expandafter \@secondoftwo
 \fi
}%
\providecommand \natexlab [1]{#1}%
\providecommand \enquote  [1]{``#1''}%
\providecommand \bibnamefont  [1]{#1}%
\providecommand \bibfnamefont [1]{#1}%
\providecommand \citenamefont [1]{#1}%
\providecommand \href@noop [0]{\@secondoftwo}%
\providecommand \href [0]{\begingroup \@sanitize@url \@href}%
\providecommand \@href[1]{\@@startlink{#1}\@@href}%
\providecommand \@@href[1]{\endgroup#1\@@endlink}%
\providecommand \@sanitize@url [0]{\catcode `\\12\catcode `\$12\catcode
  `\&12\catcode `\#12\catcode `\^12\catcode `\_12\catcode `\%12\relax}%
\providecommand \@@startlink[1]{}%
\providecommand \@@endlink[0]{}%
\providecommand \url  [0]{\begingroup\@sanitize@url \@url }%
\providecommand \@url [1]{\endgroup\@href {#1}{\urlprefix }}%
\providecommand \urlprefix  [0]{URL }%
\providecommand \Eprint [0]{\href }%
\providecommand \doibase [0]{https://doi.org/}%
\providecommand \selectlanguage [0]{\@gobble}%
\providecommand \bibinfo  [0]{\@secondoftwo}%
\providecommand \bibfield  [0]{\@secondoftwo}%
\providecommand \translation [1]{[#1]}%
\providecommand \BibitemOpen [0]{}%
\providecommand \bibitemStop [0]{}%
\providecommand \bibitemNoStop [0]{.\EOS\space}%
\providecommand \EOS [0]{\spacefactor3000\relax}%
\providecommand \BibitemShut  [1]{\csname bibitem#1\endcsname}%
\let\auto@bib@innerbib\@empty
%</preamble>
\bibitem [{\citenamefont {DiVincenzo}(2000)}]{DiVincenzo2000-ei}%
  \BibitemOpen
  \bibfield  {author} {\bibinfo {author} {\bibfnamefont {D.~P.}\ \bibnamefont
  {DiVincenzo}},\ }\bibfield  {title} {\bibinfo {title} {The physical
  implementation of quantum computation},\ }\href
  {https://doi.org/10.1002/1521-3978(200009)48:9/11<771::aid-prop771>3.0.co;2-e}
  {\bibfield  {journal} {\bibinfo  {journal} {Fortschr. Phys.}\ }\textbf
  {\bibinfo {volume} {48}},\ \bibinfo {pages} {771} (\bibinfo {year}
  {2000})}\BibitemShut {NoStop}%
\bibitem [{\citenamefont {Shor}(1995)}]{Shor1995}%
  \BibitemOpen
  \bibfield  {author} {\bibinfo {author} {\bibfnamefont {P.~W.}\ \bibnamefont
  {Shor}},\ }\bibfield  {title} {\bibinfo {title} {Scheme for reducing
  decoherence in quantum computer memory},\ }\href
  {https://doi.org/10.1103/PhysRevA.52.R2493} {\bibfield  {journal} {\bibinfo
  {journal} {Phys. Rev. A}\ }\textbf {\bibinfo {volume} {52}},\ \bibinfo
  {pages} {R2493} (\bibinfo {year} {1995})}\BibitemShut {NoStop}%
\bibitem [{\citenamefont {Knill}(2005)}]{knill2005quantum}%
  \BibitemOpen
  \bibfield  {author} {\bibinfo {author} {\bibfnamefont {E.}~\bibnamefont
  {Knill}},\ }\bibfield  {title} {\bibinfo {title} {Quantum computing with
  realistically noisy devices},\ }\href
  {https://www.nature.com/articles/nature03350} {\bibfield  {journal} {\bibinfo
   {journal} {Nature}\ }\textbf {\bibinfo {volume} {434}},\ \bibinfo {pages}
  {39} (\bibinfo {year} {2005})}\BibitemShut {NoStop}%
\bibitem [{\citenamefont {DiVincenzo}\ and\ \citenamefont
  {Aliferis}(2007)}]{divincenzo2007effective}%
  \BibitemOpen
  \bibfield  {author} {\bibinfo {author} {\bibfnamefont {D.~P.}\ \bibnamefont
  {DiVincenzo}}\ and\ \bibinfo {author} {\bibfnamefont {P.}~\bibnamefont
  {Aliferis}},\ }\bibfield  {title} {\bibinfo {title} {Effective fault-tolerant
  quantum computation with slow measurements},\ }\href@noop {} {\bibfield
  {journal} {\bibinfo  {journal} {Physical review letters}\ }\textbf {\bibinfo
  {volume} {98}},\ \bibinfo {pages} {020501} (\bibinfo {year}
  {2007})}\BibitemShut {NoStop}%
\bibitem [{\citenamefont {Landahl}\ \emph {et~al.}(2011)\citenamefont
  {Landahl}, \citenamefont {Anderson},\ and\ \citenamefont
  {Rice}}]{landahl2011fault}%
  \BibitemOpen
  \bibfield  {author} {\bibinfo {author} {\bibfnamefont {A.~J.}\ \bibnamefont
  {Landahl}}, \bibinfo {author} {\bibfnamefont {J.~T.}\ \bibnamefont
  {Anderson}},\ and\ \bibinfo {author} {\bibfnamefont {P.~R.}\ \bibnamefont
  {Rice}},\ }\bibfield  {title} {\bibinfo {title} {Fault-tolerant quantum
  computing with color codes},\ }\href@noop {} {\bibfield  {journal} {\bibinfo
  {journal} {arXiv preprint arXiv:1108.5738}\ } (\bibinfo {year}
  {2011})}\BibitemShut {NoStop}%
\bibitem [{\citenamefont {Fowler}\ \emph {et~al.}(2012)\citenamefont {Fowler},
  \citenamefont {Mariantoni}, \citenamefont {Martinis},\ and\ \citenamefont
  {Cleland}}]{Fowler2012-ea}%
  \BibitemOpen
  \bibfield  {author} {\bibinfo {author} {\bibfnamefont {A.~G.}\ \bibnamefont
  {Fowler}}, \bibinfo {author} {\bibfnamefont {M.}~\bibnamefont {Mariantoni}},
  \bibinfo {author} {\bibfnamefont {J.~M.}\ \bibnamefont {Martinis}},\ and\
  \bibinfo {author} {\bibfnamefont {A.~N.}\ \bibnamefont {Cleland}},\
  }\bibfield  {title} {\bibinfo {title} {Surface codes: Towards practical
  large-scale quantum computation},\ }\href
  {https://doi.org/10.1103/PhysRevA.86.032324} {\bibfield  {journal} {\bibinfo
  {journal} {Phys. Rev. A}\ }\textbf {\bibinfo {volume} {86}},\ \bibinfo
  {pages} {032324} (\bibinfo {year} {2012})}\BibitemShut {NoStop}%
\bibitem [{\citenamefont {Devitt}\ \emph {et~al.}(2013)\citenamefont {Devitt},
  \citenamefont {Munro},\ and\ \citenamefont {Nemoto}}]{devitt2013quantum}%
  \BibitemOpen
  \bibfield  {author} {\bibinfo {author} {\bibfnamefont {S.~J.}\ \bibnamefont
  {Devitt}}, \bibinfo {author} {\bibfnamefont {W.~J.}\ \bibnamefont {Munro}},\
  and\ \bibinfo {author} {\bibfnamefont {K.}~\bibnamefont {Nemoto}},\
  }\bibfield  {title} {\bibinfo {title} {Quantum error correction for
  beginners},\ }\href@noop {} {\bibfield  {journal} {\bibinfo  {journal}
  {Reports on Progress in Physics}\ }\textbf {\bibinfo {volume} {76}},\
  \bibinfo {pages} {076001} (\bibinfo {year} {2013})}\BibitemShut {NoStop}%
\bibitem [{\citenamefont {Rist{\`e}}\ \emph {et~al.}(2020)\citenamefont
  {Rist{\`e}}, \citenamefont {Govia}, \citenamefont {Donovan}, \citenamefont
  {Fallek}, \citenamefont {Kalfus}, \citenamefont {Brink}, \citenamefont
  {Bronn},\ and\ \citenamefont {Ohki}}]{Riste2020}%
  \BibitemOpen
  \bibfield  {author} {\bibinfo {author} {\bibfnamefont {D.}~\bibnamefont
  {Rist{\`e}}}, \bibinfo {author} {\bibfnamefont {L.~C.~G.}\ \bibnamefont
  {Govia}}, \bibinfo {author} {\bibfnamefont {B.}~\bibnamefont {Donovan}},
  \bibinfo {author} {\bibfnamefont {S.~D.}\ \bibnamefont {Fallek}}, \bibinfo
  {author} {\bibfnamefont {W.~D.}\ \bibnamefont {Kalfus}}, \bibinfo {author}
  {\bibfnamefont {M.}~\bibnamefont {Brink}}, \bibinfo {author} {\bibfnamefont
  {N.~T.}\ \bibnamefont {Bronn}},\ and\ \bibinfo {author} {\bibfnamefont
  {T.~A.}\ \bibnamefont {Ohki}},\ }\bibfield  {title} {\bibinfo {title}
  {Real-time processing of stabilizer measurements in a bit-flip code},\ }\href
  {https://doi.org/10.1038/s41534-020-00304-y} {\bibfield  {journal} {\bibinfo
  {journal} {npj Quantum Information}\ }\textbf {\bibinfo {volume} {6}},\
  \bibinfo {pages} {71} (\bibinfo {year} {2020})}\BibitemShut {NoStop}%
\bibitem [{\citenamefont {Corcoles}\ \emph {et~al.}(2021)\citenamefont
  {Corcoles}, \citenamefont {Takita}, \citenamefont {Inoue}, \citenamefont
  {Lekuch}, \citenamefont {Minev}, \citenamefont {Chow},\ and\ \citenamefont
  {Gambetta}}]{Corcoles2021-ht}%
  \BibitemOpen
  \bibfield  {author} {\bibinfo {author} {\bibfnamefont {A.~D.}\ \bibnamefont
  {Corcoles}}, \bibinfo {author} {\bibfnamefont {M.}~\bibnamefont {Takita}},
  \bibinfo {author} {\bibfnamefont {K.}~\bibnamefont {Inoue}}, \bibinfo
  {author} {\bibfnamefont {S.}~\bibnamefont {Lekuch}}, \bibinfo {author}
  {\bibfnamefont {Z.~K.}\ \bibnamefont {Minev}}, \bibinfo {author}
  {\bibfnamefont {J.~M.}\ \bibnamefont {Chow}},\ and\ \bibinfo {author}
  {\bibfnamefont {J.~M.}\ \bibnamefont {Gambetta}},\ }\bibfield  {title}
  {\bibinfo {title} {Exploiting dynamic quantum circuits in a quantum algorithm
  with superconducting qubits},\ }\href {http://arxiv.org/abs/2102.01682}
  {\bibfield  {journal} {\bibinfo  {journal} {arxiv}\ } (\bibinfo {year}
  {2021})},\ \Eprint {https://arxiv.org/abs/2102.01682} {arXiv:2102.01682
  [quant-ph]} \BibitemShut {NoStop}%
\bibitem [{\citenamefont {Urbanek}\ \emph {et~al.}(2020)\citenamefont
  {Urbanek}, \citenamefont {Nachman},\ and\ \citenamefont
  {de~Jong}}]{Urbanek2020-nm}%
  \BibitemOpen
  \bibfield  {author} {\bibinfo {author} {\bibfnamefont {M.}~\bibnamefont
  {Urbanek}}, \bibinfo {author} {\bibfnamefont {B.}~\bibnamefont {Nachman}},\
  and\ \bibinfo {author} {\bibfnamefont {W.~A.}\ \bibnamefont {de~Jong}},\
  }\bibfield  {title} {\bibinfo {title} {Error detection on quantum computers
  improving the accuracy of chemical calculations},\ }\href
  {https://doi.org/10.1103/PhysRevA.102.022427} {\bibfield  {journal} {\bibinfo
   {journal} {Phys. Rev. A}\ }\textbf {\bibinfo {volume} {102}},\ \bibinfo
  {pages} {022427} (\bibinfo {year} {2020})}\BibitemShut {NoStop}%
\bibitem [{\citenamefont {Holmes}\ \emph {et~al.}(2020)\citenamefont {Holmes},
  \citenamefont {Jokar}, \citenamefont {Pasandi}, \citenamefont {Ding},
  \citenamefont {Pedram},\ and\ \citenamefont {Chong}}]{Holmes2020-wl}%
  \BibitemOpen
  \bibfield  {author} {\bibinfo {author} {\bibfnamefont {A.}~\bibnamefont
  {Holmes}}, \bibinfo {author} {\bibfnamefont {M.~R.}\ \bibnamefont {Jokar}},
  \bibinfo {author} {\bibfnamefont {G.}~\bibnamefont {Pasandi}}, \bibinfo
  {author} {\bibfnamefont {Y.}~\bibnamefont {Ding}}, \bibinfo {author}
  {\bibfnamefont {M.}~\bibnamefont {Pedram}},\ and\ \bibinfo {author}
  {\bibfnamefont {F.~T.}\ \bibnamefont {Chong}},\ }\bibfield  {title} {\bibinfo
  {title} {{NISQ+}: Boosting quantum computing power by approximating quantum
  error correction},\ }in\ \href {https://doi.org/10.1109/ISCA45697.2020.00053}
  {\emph {\bibinfo {booktitle} {2020 {ACM/IEEE} 47th Annual International
  Symposium on Computer Architecture ({ISCA})}}}\ (\bibinfo {year} {2020})\
  pp.\ \bibinfo {pages} {556--569}\BibitemShut {NoStop}%
\bibitem [{\citenamefont {Paetznick}\ and\ \citenamefont
  {Svore}(2013)}]{paetznick2013repeat}%
  \BibitemOpen
  \bibfield  {author} {\bibinfo {author} {\bibfnamefont {A.}~\bibnamefont
  {Paetznick}}\ and\ \bibinfo {author} {\bibfnamefont {K.~M.}\ \bibnamefont
  {Svore}},\ }\bibfield  {title} {\bibinfo {title} {Repeat-until-success:
  Non-deterministic decomposition of single-qubit unitaries},\ }\href@noop {}
  {\bibfield  {journal} {\bibinfo  {journal} {arXiv preprint arXiv:1311.1074}\
  } (\bibinfo {year} {2013})}\BibitemShut {NoStop}%
\bibitem [{\citenamefont {Greenbaum}(2015)}]{Greenbaum2015-tr}%
  \BibitemOpen
  \bibfield  {author} {\bibinfo {author} {\bibfnamefont {D.}~\bibnamefont
  {Greenbaum}},\ }\bibfield  {title} {\bibinfo {title} {Introduction to quantum
  gate set tomography},\ }\href@noop {} {\bibfield  {journal} {\bibinfo
  {journal} {arxiv}\ } (\bibinfo {year} {2015})},\ \Eprint
  {https://arxiv.org/abs/1509.02921} {arXiv:1509.02921 [quant-ph]} \BibitemShut
  {NoStop}%
\bibitem [{\citenamefont {Blume-Kohout}\ \emph {et~al.}(2017)\citenamefont
  {Blume-Kohout}, \citenamefont {Gamble}, \citenamefont {Nielsen},
  \citenamefont {Rudinger}, \citenamefont {Mizrahi}, \citenamefont {Fortier},\
  and\ \citenamefont {Maunz}}]{Blume-Kohout2017-no}%
  \BibitemOpen
  \bibfield  {author} {\bibinfo {author} {\bibfnamefont {R.}~\bibnamefont
  {Blume-Kohout}}, \bibinfo {author} {\bibfnamefont {J.~K.}\ \bibnamefont
  {Gamble}}, \bibinfo {author} {\bibfnamefont {E.}~\bibnamefont {Nielsen}},
  \bibinfo {author} {\bibfnamefont {K.}~\bibnamefont {Rudinger}}, \bibinfo
  {author} {\bibfnamefont {J.}~\bibnamefont {Mizrahi}}, \bibinfo {author}
  {\bibfnamefont {K.}~\bibnamefont {Fortier}},\ and\ \bibinfo {author}
  {\bibfnamefont {P.}~\bibnamefont {Maunz}},\ }\bibfield  {title} {\bibinfo
  {title} {Demonstration of qubit operations below a rigorous fault tolerance
  threshold with gate set tomography},\ }\href
  {https://doi.org/10.1038/ncomms14485} {\bibfield  {journal} {\bibinfo
  {journal} {Nat. Commun.}\ }\textbf {\bibinfo {volume} {8}},\ \bibinfo {pages}
  {14485} (\bibinfo {year} {2017})}\BibitemShut {NoStop}%
\bibitem [{\citenamefont {Nielsen}\ \emph
  {et~al.}(2020{\natexlab{a}})\citenamefont {Nielsen}, \citenamefont {Gamble},
  \citenamefont {Rudinger}, \citenamefont {Scholten}, \citenamefont {Young},\
  and\ \citenamefont {Blume-Kohout}}]{Nielsen2020-lt}%
  \BibitemOpen
  \bibfield  {author} {\bibinfo {author} {\bibfnamefont {E.}~\bibnamefont
  {Nielsen}}, \bibinfo {author} {\bibfnamefont {J.~K.}\ \bibnamefont {Gamble}},
  \bibinfo {author} {\bibfnamefont {K.}~\bibnamefont {Rudinger}}, \bibinfo
  {author} {\bibfnamefont {T.}~\bibnamefont {Scholten}}, \bibinfo {author}
  {\bibfnamefont {K.}~\bibnamefont {Young}},\ and\ \bibinfo {author}
  {\bibfnamefont {R.}~\bibnamefont {Blume-Kohout}},\ }\bibfield  {title}
  {\bibinfo {title} {Gate set tomography},\ }\href
  {http://arxiv.org/abs/2009.07301} {\bibfield  {journal} {\bibinfo  {journal}
  {arxiv}\ } (\bibinfo {year} {2020}{\natexlab{a}})},\ \Eprint
  {https://arxiv.org/abs/2009.07301} {arXiv:2009.07301 [quant-ph]} \BibitemShut
  {NoStop}%
\bibitem [{\citenamefont {Nielsen}\ \emph
  {et~al.}(2020{\natexlab{b}})\citenamefont {Nielsen}, \citenamefont
  {Rudinger}, \citenamefont {Proctor}, \citenamefont {Russo}, \citenamefont
  {Young},\ and\ \citenamefont {Blume-Kohout}}]{Nielsen2020-lu}%
  \BibitemOpen
  \bibfield  {author} {\bibinfo {author} {\bibfnamefont {E.}~\bibnamefont
  {Nielsen}}, \bibinfo {author} {\bibfnamefont {K.}~\bibnamefont {Rudinger}},
  \bibinfo {author} {\bibfnamefont {T.}~\bibnamefont {Proctor}}, \bibinfo
  {author} {\bibfnamefont {A.}~\bibnamefont {Russo}}, \bibinfo {author}
  {\bibfnamefont {K.}~\bibnamefont {Young}},\ and\ \bibinfo {author}
  {\bibfnamefont {R.}~\bibnamefont {Blume-Kohout}},\ }\bibfield  {title}
  {\bibinfo {title} {Probing quantum processor performance with {pyGSTi}},\
  }\href@noop {} {\bibfield  {journal} {\bibinfo  {journal} {arXiv}\ }
  (\bibinfo {year} {2020}{\natexlab{b}})},\ \Eprint
  {https://arxiv.org/abs/2002.12476} {arXiv:2002.12476 [quant-ph]} \BibitemShut
  {NoStop}%
\bibitem [{\citenamefont {Chen}\ \emph {et~al.}(2021)\citenamefont {Chen},
  \citenamefont {Satzinger}, \citenamefont {Atalaya}, \citenamefont {Korotkov},
  \citenamefont {Dunsworth}, \citenamefont {Sank}, \citenamefont {Quintana},
  \citenamefont {McEwen}, \citenamefont {Barends}, \citenamefont {Klimov},
  \citenamefont {Hong}, \citenamefont {Jones}, \citenamefont {Petukhov},
  \citenamefont {Kafri}, \citenamefont {Demura}, \citenamefont {Burkett},
  \citenamefont {Gidney}, \citenamefont {Fowler}, \citenamefont {Putterman},
  \citenamefont {Aleiner}, \citenamefont {Arute}, \citenamefont {Arya},
  \citenamefont {Babbush}, \citenamefont {Bardin}, \citenamefont {Bengtsson},
  \citenamefont {Bourassa}, \citenamefont {Broughton}, \citenamefont {Buckley},
  \citenamefont {Buell}, \citenamefont {Bushnell}, \citenamefont {Chiaro},
  \citenamefont {Collins}, \citenamefont {Courtney}, \citenamefont {Derk},
  \citenamefont {Eppens}, \citenamefont {Erickson}, \citenamefont {Farhi},
  \citenamefont {Foxen}, \citenamefont {Giustina}, \citenamefont {Gross},
  \citenamefont {Harrigan}, \citenamefont {Harrington}, \citenamefont {Hilton},
  \citenamefont {Ho}, \citenamefont {Huang}, \citenamefont {Huggins},
  \citenamefont {Ioffe}, \citenamefont {Isakov}, \citenamefont {Jeffrey},
  \citenamefont {Jiang}, \citenamefont {Kechedzhi}, \citenamefont {Kim},
  \citenamefont {Kostritsa}, \citenamefont {Landhuis}, \citenamefont {Laptev},
  \citenamefont {Lucero}, \citenamefont {Martin}, \citenamefont {McClean},
  \citenamefont {McCourt}, \citenamefont {Mi}, \citenamefont {Miao},
  \citenamefont {Mohseni}, \citenamefont {Mruczkiewicz}, \citenamefont {Mutus},
  \citenamefont {Naaman}, \citenamefont {Neeley}, \citenamefont {Neill},
  \citenamefont {Newman}, \citenamefont {Niu}, \citenamefont {O'Brien},
  \citenamefont {Opremcak}, \citenamefont {Ostby}, \citenamefont {Pat{\'o}},
  \citenamefont {Redd}, \citenamefont {Roushan}, \citenamefont {Rubin},
  \citenamefont {Shvarts}, \citenamefont {Strain}, \citenamefont {Szalay},
  \citenamefont {Trevithick}, \citenamefont {Villalonga}, \citenamefont
  {White}, \citenamefont {Jamie~Yao}, \citenamefont {Yeh}, \citenamefont
  {Zalcman}, \citenamefont {Neven}, \citenamefont {Boixo}, \citenamefont
  {Smelyanskiy}, \citenamefont {Chen}, \citenamefont {Megrant},\ and\
  \citenamefont {Kelly}}]{Chen2021-nv}%
  \BibitemOpen
  \bibfield  {author} {\bibinfo {author} {\bibfnamefont {Z.}~\bibnamefont
  {Chen}}, \bibinfo {author} {\bibfnamefont {K.~J.}\ \bibnamefont {Satzinger}},
  \bibinfo {author} {\bibfnamefont {J.}~\bibnamefont {Atalaya}}, \bibinfo
  {author} {\bibfnamefont {A.~N.}\ \bibnamefont {Korotkov}}, \bibinfo {author}
  {\bibfnamefont {A.}~\bibnamefont {Dunsworth}}, \bibinfo {author}
  {\bibfnamefont {D.}~\bibnamefont {Sank}}, \bibinfo {author} {\bibfnamefont
  {C.}~\bibnamefont {Quintana}}, \bibinfo {author} {\bibfnamefont
  {M.}~\bibnamefont {McEwen}}, \bibinfo {author} {\bibfnamefont
  {R.}~\bibnamefont {Barends}}, \bibinfo {author} {\bibfnamefont {P.~V.}\
  \bibnamefont {Klimov}}, \bibinfo {author} {\bibfnamefont {S.}~\bibnamefont
  {Hong}}, \bibinfo {author} {\bibfnamefont {C.}~\bibnamefont {Jones}},
  \bibinfo {author} {\bibfnamefont {A.}~\bibnamefont {Petukhov}}, \bibinfo
  {author} {\bibfnamefont {D.}~\bibnamefont {Kafri}}, \bibinfo {author}
  {\bibfnamefont {S.}~\bibnamefont {Demura}}, \bibinfo {author} {\bibfnamefont
  {B.}~\bibnamefont {Burkett}}, \bibinfo {author} {\bibfnamefont
  {C.}~\bibnamefont {Gidney}}, \bibinfo {author} {\bibfnamefont {A.~G.}\
  \bibnamefont {Fowler}}, \bibinfo {author} {\bibfnamefont {H.}~\bibnamefont
  {Putterman}}, \bibinfo {author} {\bibfnamefont {I.}~\bibnamefont {Aleiner}},
  \bibinfo {author} {\bibfnamefont {F.}~\bibnamefont {Arute}}, \bibinfo
  {author} {\bibfnamefont {K.}~\bibnamefont {Arya}}, \bibinfo {author}
  {\bibfnamefont {R.}~\bibnamefont {Babbush}}, \bibinfo {author} {\bibfnamefont
  {J.~C.}\ \bibnamefont {Bardin}}, \bibinfo {author} {\bibfnamefont
  {A.}~\bibnamefont {Bengtsson}}, \bibinfo {author} {\bibfnamefont
  {A.}~\bibnamefont {Bourassa}}, \bibinfo {author} {\bibfnamefont
  {M.}~\bibnamefont {Broughton}}, \bibinfo {author} {\bibfnamefont {B.~B.}\
  \bibnamefont {Buckley}}, \bibinfo {author} {\bibfnamefont {D.~A.}\
  \bibnamefont {Buell}}, \bibinfo {author} {\bibfnamefont {N.}~\bibnamefont
  {Bushnell}}, \bibinfo {author} {\bibfnamefont {B.}~\bibnamefont {Chiaro}},
  \bibinfo {author} {\bibfnamefont {R.}~\bibnamefont {Collins}}, \bibinfo
  {author} {\bibfnamefont {W.}~\bibnamefont {Courtney}}, \bibinfo {author}
  {\bibfnamefont {A.~R.}\ \bibnamefont {Derk}}, \bibinfo {author}
  {\bibfnamefont {D.}~\bibnamefont {Eppens}}, \bibinfo {author} {\bibfnamefont
  {C.}~\bibnamefont {Erickson}}, \bibinfo {author} {\bibfnamefont
  {E.}~\bibnamefont {Farhi}}, \bibinfo {author} {\bibfnamefont
  {B.}~\bibnamefont {Foxen}}, \bibinfo {author} {\bibfnamefont
  {M.}~\bibnamefont {Giustina}}, \bibinfo {author} {\bibfnamefont {J.~A.}\
  \bibnamefont {Gross}}, \bibinfo {author} {\bibfnamefont {M.~P.}\ \bibnamefont
  {Harrigan}}, \bibinfo {author} {\bibfnamefont {S.~D.}\ \bibnamefont
  {Harrington}}, \bibinfo {author} {\bibfnamefont {J.}~\bibnamefont {Hilton}},
  \bibinfo {author} {\bibfnamefont {A.}~\bibnamefont {Ho}}, \bibinfo {author}
  {\bibfnamefont {T.}~\bibnamefont {Huang}}, \bibinfo {author} {\bibfnamefont
  {W.~J.}\ \bibnamefont {Huggins}}, \bibinfo {author} {\bibfnamefont {L.~B.}\
  \bibnamefont {Ioffe}}, \bibinfo {author} {\bibfnamefont {S.~V.}\ \bibnamefont
  {Isakov}}, \bibinfo {author} {\bibfnamefont {E.}~\bibnamefont {Jeffrey}},
  \bibinfo {author} {\bibfnamefont {Z.}~\bibnamefont {Jiang}}, \bibinfo
  {author} {\bibfnamefont {K.}~\bibnamefont {Kechedzhi}}, \bibinfo {author}
  {\bibfnamefont {S.}~\bibnamefont {Kim}}, \bibinfo {author} {\bibfnamefont
  {F.}~\bibnamefont {Kostritsa}}, \bibinfo {author} {\bibfnamefont
  {D.}~\bibnamefont {Landhuis}}, \bibinfo {author} {\bibfnamefont
  {P.}~\bibnamefont {Laptev}}, \bibinfo {author} {\bibfnamefont
  {E.}~\bibnamefont {Lucero}}, \bibinfo {author} {\bibfnamefont
  {O.}~\bibnamefont {Martin}}, \bibinfo {author} {\bibfnamefont {J.~R.}\
  \bibnamefont {McClean}}, \bibinfo {author} {\bibfnamefont {T.}~\bibnamefont
  {McCourt}}, \bibinfo {author} {\bibfnamefont {X.}~\bibnamefont {Mi}},
  \bibinfo {author} {\bibfnamefont {K.~C.}\ \bibnamefont {Miao}}, \bibinfo
  {author} {\bibfnamefont {M.}~\bibnamefont {Mohseni}}, \bibinfo {author}
  {\bibfnamefont {W.}~\bibnamefont {Mruczkiewicz}}, \bibinfo {author}
  {\bibfnamefont {J.}~\bibnamefont {Mutus}}, \bibinfo {author} {\bibfnamefont
  {O.}~\bibnamefont {Naaman}}, \bibinfo {author} {\bibfnamefont
  {M.}~\bibnamefont {Neeley}}, \bibinfo {author} {\bibfnamefont
  {C.}~\bibnamefont {Neill}}, \bibinfo {author} {\bibfnamefont
  {M.}~\bibnamefont {Newman}}, \bibinfo {author} {\bibfnamefont {M.~Y.}\
  \bibnamefont {Niu}}, \bibinfo {author} {\bibfnamefont {T.~E.}\ \bibnamefont
  {O'Brien}}, \bibinfo {author} {\bibfnamefont {A.}~\bibnamefont {Opremcak}},
  \bibinfo {author} {\bibfnamefont {E.}~\bibnamefont {Ostby}}, \bibinfo
  {author} {\bibfnamefont {B.}~\bibnamefont {Pat{\'o}}}, \bibinfo {author}
  {\bibfnamefont {N.}~\bibnamefont {Redd}}, \bibinfo {author} {\bibfnamefont
  {P.}~\bibnamefont {Roushan}}, \bibinfo {author} {\bibfnamefont {N.~C.}\
  \bibnamefont {Rubin}}, \bibinfo {author} {\bibfnamefont {V.}~\bibnamefont
  {Shvarts}}, \bibinfo {author} {\bibfnamefont {D.}~\bibnamefont {Strain}},
  \bibinfo {author} {\bibfnamefont {M.}~\bibnamefont {Szalay}}, \bibinfo
  {author} {\bibfnamefont {M.~D.}\ \bibnamefont {Trevithick}}, \bibinfo
  {author} {\bibfnamefont {B.}~\bibnamefont {Villalonga}}, \bibinfo {author}
  {\bibfnamefont {T.}~\bibnamefont {White}}, \bibinfo {author} {\bibfnamefont
  {Z.}~\bibnamefont {Jamie~Yao}}, \bibinfo {author} {\bibfnamefont
  {P.}~\bibnamefont {Yeh}}, \bibinfo {author} {\bibfnamefont {A.}~\bibnamefont
  {Zalcman}}, \bibinfo {author} {\bibfnamefont {H.}~\bibnamefont {Neven}},
  \bibinfo {author} {\bibfnamefont {S.}~\bibnamefont {Boixo}}, \bibinfo
  {author} {\bibfnamefont {V.}~\bibnamefont {Smelyanskiy}}, \bibinfo {author}
  {\bibfnamefont {Y.}~\bibnamefont {Chen}}, \bibinfo {author} {\bibfnamefont
  {A.}~\bibnamefont {Megrant}},\ and\ \bibinfo {author} {\bibfnamefont
  {J.}~\bibnamefont {Kelly}},\ }\bibfield  {title} {\bibinfo {title}
  {Exponential suppression of bit or phase flip errors with repetitive error
  correction},\ }\href {http://arxiv.org/abs/2102.06132} {\  (\bibinfo {year}
  {2021})},\ \Eprint {https://arxiv.org/abs/2102.06132} {arXiv:2102.06132
  [quant-ph]} \BibitemShut {NoStop}%
\bibitem [{\citenamefont {Andersen}\ \emph {et~al.}(2020)\citenamefont
  {Andersen}, \citenamefont {Remm}, \citenamefont {Lazar}, \citenamefont
  {Krinner}, \citenamefont {Lacroix}, \citenamefont {Norris}, \citenamefont
  {Gabureac}, \citenamefont {Eichler},\ and\ \citenamefont
  {Wallraff}}]{andersen2020repeated}%
  \BibitemOpen
  \bibfield  {author} {\bibinfo {author} {\bibfnamefont {C.~K.}\ \bibnamefont
  {Andersen}}, \bibinfo {author} {\bibfnamefont {A.}~\bibnamefont {Remm}},
  \bibinfo {author} {\bibfnamefont {S.}~\bibnamefont {Lazar}}, \bibinfo
  {author} {\bibfnamefont {S.}~\bibnamefont {Krinner}}, \bibinfo {author}
  {\bibfnamefont {N.}~\bibnamefont {Lacroix}}, \bibinfo {author} {\bibfnamefont
  {G.~J.}\ \bibnamefont {Norris}}, \bibinfo {author} {\bibfnamefont
  {M.}~\bibnamefont {Gabureac}}, \bibinfo {author} {\bibfnamefont
  {C.}~\bibnamefont {Eichler}},\ and\ \bibinfo {author} {\bibfnamefont
  {A.}~\bibnamefont {Wallraff}},\ }\bibfield  {title} {\bibinfo {title}
  {Repeated quantum error detection in a surface code},\ }\href@noop {}
  {\bibfield  {journal} {\bibinfo  {journal} {Nature Physics}\ }\textbf
  {\bibinfo {volume} {16}},\ \bibinfo {pages} {875} (\bibinfo {year}
  {2020})}\BibitemShut {NoStop}%
\bibitem [{\citenamefont {Erhard}\ \emph {et~al.}(2021)\citenamefont {Erhard},
  \citenamefont {Nautrup}, \citenamefont {Meth}, \citenamefont {Postler},
  \citenamefont {Stricker}, \citenamefont {Stadler}, \citenamefont
  {Negnevitsky}, \citenamefont {Ringbauer}, \citenamefont {Schindler},
  \citenamefont {Briegel} \emph {et~al.}}]{erhard2021entangling}%
  \BibitemOpen
  \bibfield  {author} {\bibinfo {author} {\bibfnamefont {A.}~\bibnamefont
  {Erhard}}, \bibinfo {author} {\bibfnamefont {H.~P.}\ \bibnamefont {Nautrup}},
  \bibinfo {author} {\bibfnamefont {M.}~\bibnamefont {Meth}}, \bibinfo {author}
  {\bibfnamefont {L.}~\bibnamefont {Postler}}, \bibinfo {author} {\bibfnamefont
  {R.}~\bibnamefont {Stricker}}, \bibinfo {author} {\bibfnamefont
  {M.}~\bibnamefont {Stadler}}, \bibinfo {author} {\bibfnamefont
  {V.}~\bibnamefont {Negnevitsky}}, \bibinfo {author} {\bibfnamefont
  {M.}~\bibnamefont {Ringbauer}}, \bibinfo {author} {\bibfnamefont
  {P.}~\bibnamefont {Schindler}}, \bibinfo {author} {\bibfnamefont {H.~J.}\
  \bibnamefont {Briegel}}, \emph {et~al.},\ }\bibfield  {title} {\bibinfo
  {title} {Entangling logical qubits with lattice surgery},\ }\href@noop {}
  {\bibfield  {journal} {\bibinfo  {journal} {Nature}\ }\textbf {\bibinfo
  {volume} {589}},\ \bibinfo {pages} {220} (\bibinfo {year}
  {2021})}\BibitemShut {NoStop}%
\bibitem [{\citenamefont {Negnevitsky}\ \emph {et~al.}(2018)\citenamefont
  {Negnevitsky}, \citenamefont {Marinelli}, \citenamefont {Mehta},
  \citenamefont {Lo}, \citenamefont {Fl{\"u}hmann},\ and\ \citenamefont
  {Home}}]{negnevitsky2018repeated}%
  \BibitemOpen
  \bibfield  {author} {\bibinfo {author} {\bibfnamefont {V.}~\bibnamefont
  {Negnevitsky}}, \bibinfo {author} {\bibfnamefont {M.}~\bibnamefont
  {Marinelli}}, \bibinfo {author} {\bibfnamefont {K.~K.}\ \bibnamefont
  {Mehta}}, \bibinfo {author} {\bibfnamefont {H.-Y.}\ \bibnamefont {Lo}},
  \bibinfo {author} {\bibfnamefont {C.}~\bibnamefont {Fl{\"u}hmann}},\ and\
  \bibinfo {author} {\bibfnamefont {J.~P.}\ \bibnamefont {Home}},\ }\bibfield
  {title} {\bibinfo {title} {Repeated multi-qubit readout and feedback with a
  mixed-species trapped-ion register},\ }\href@noop {} {\bibfield  {journal}
  {\bibinfo  {journal} {Nature}\ }\textbf {\bibinfo {volume} {563}},\ \bibinfo
  {pages} {527} (\bibinfo {year} {2018})}\BibitemShut {NoStop}%
\bibitem [{\citenamefont {Rosenblum}\ \emph {et~al.}(2018)\citenamefont
  {Rosenblum}, \citenamefont {Reinhold}, \citenamefont {Mirrahimi},
  \citenamefont {Jiang}, \citenamefont {Frunzio},\ and\ \citenamefont
  {Schoelkopf}}]{rosenblum2018fault}%
  \BibitemOpen
  \bibfield  {author} {\bibinfo {author} {\bibfnamefont {S.}~\bibnamefont
  {Rosenblum}}, \bibinfo {author} {\bibfnamefont {P.}~\bibnamefont {Reinhold}},
  \bibinfo {author} {\bibfnamefont {M.}~\bibnamefont {Mirrahimi}}, \bibinfo
  {author} {\bibfnamefont {L.}~\bibnamefont {Jiang}}, \bibinfo {author}
  {\bibfnamefont {L.}~\bibnamefont {Frunzio}},\ and\ \bibinfo {author}
  {\bibfnamefont {R.~J.}\ \bibnamefont {Schoelkopf}},\ }\bibfield  {title}
  {\bibinfo {title} {Fault-tolerant detection of a quantum error},\ }\href@noop
  {} {\bibfield  {journal} {\bibinfo  {journal} {Science}\ }\textbf {\bibinfo
  {volume} {361}},\ \bibinfo {pages} {266} (\bibinfo {year}
  {2018})}\BibitemShut {NoStop}%
\bibitem [{\citenamefont {Elder}\ \emph {et~al.}(2020)\citenamefont {Elder},
  \citenamefont {Wang}, \citenamefont {Reinhold}, \citenamefont {Hann},
  \citenamefont {Chou}, \citenamefont {Lester}, \citenamefont {Rosenblum},
  \citenamefont {Frunzio}, \citenamefont {Jiang},\ and\ \citenamefont
  {Schoelkopf}}]{elder2020high}%
  \BibitemOpen
  \bibfield  {author} {\bibinfo {author} {\bibfnamefont {S.~S.}\ \bibnamefont
  {Elder}}, \bibinfo {author} {\bibfnamefont {C.~S.}\ \bibnamefont {Wang}},
  \bibinfo {author} {\bibfnamefont {P.}~\bibnamefont {Reinhold}}, \bibinfo
  {author} {\bibfnamefont {C.~T.}\ \bibnamefont {Hann}}, \bibinfo {author}
  {\bibfnamefont {K.~S.}\ \bibnamefont {Chou}}, \bibinfo {author}
  {\bibfnamefont {B.~J.}\ \bibnamefont {Lester}}, \bibinfo {author}
  {\bibfnamefont {S.}~\bibnamefont {Rosenblum}}, \bibinfo {author}
  {\bibfnamefont {L.}~\bibnamefont {Frunzio}}, \bibinfo {author} {\bibfnamefont
  {L.}~\bibnamefont {Jiang}},\ and\ \bibinfo {author} {\bibfnamefont {R.~J.}\
  \bibnamefont {Schoelkopf}},\ }\bibfield  {title} {\bibinfo {title}
  {High-fidelity measurement of qubits encoded in multilevel superconducting
  circuits},\ }\href@noop {} {\bibfield  {journal} {\bibinfo  {journal}
  {Physical Review X}\ }\textbf {\bibinfo {volume} {10}},\ \bibinfo {pages}
  {011001} (\bibinfo {year} {2020})}\BibitemShut {NoStop}%
\bibitem [{\citenamefont {Bultink}\ \emph {et~al.}(2020)\citenamefont
  {Bultink}, \citenamefont {O’Brien}, \citenamefont {Vollmer}, \citenamefont
  {Muthusubramanian}, \citenamefont {Beekman}, \citenamefont {Rol},
  \citenamefont {Fu}, \citenamefont {Tarasinski}, \citenamefont {Ostroukh},
  \citenamefont {Varbanov} \emph {et~al.}}]{bultink2020protecting}%
  \BibitemOpen
  \bibfield  {author} {\bibinfo {author} {\bibfnamefont {C.}~\bibnamefont
  {Bultink}}, \bibinfo {author} {\bibfnamefont {T.}~\bibnamefont {O’Brien}},
  \bibinfo {author} {\bibfnamefont {R.}~\bibnamefont {Vollmer}}, \bibinfo
  {author} {\bibfnamefont {N.}~\bibnamefont {Muthusubramanian}}, \bibinfo
  {author} {\bibfnamefont {M.}~\bibnamefont {Beekman}}, \bibinfo {author}
  {\bibfnamefont {M.}~\bibnamefont {Rol}}, \bibinfo {author} {\bibfnamefont
  {X.}~\bibnamefont {Fu}}, \bibinfo {author} {\bibfnamefont {B.}~\bibnamefont
  {Tarasinski}}, \bibinfo {author} {\bibfnamefont {V.}~\bibnamefont
  {Ostroukh}}, \bibinfo {author} {\bibfnamefont {B.}~\bibnamefont {Varbanov}},
  \emph {et~al.},\ }\bibfield  {title} {\bibinfo {title} {Protecting quantum
  entanglement from leakage and qubit errors via repetitive parity
  measurements},\ }\href@noop {} {\bibfield  {journal} {\bibinfo  {journal}
  {Science advances}\ }\textbf {\bibinfo {volume} {6}},\ \bibinfo {pages}
  {eaay3050} (\bibinfo {year} {2020})}\BibitemShut {NoStop}%
\bibitem [{\citenamefont {Fowler}\ \emph {et~al.}(2014)\citenamefont {Fowler},
  \citenamefont {Sank}, \citenamefont {Kelly}, \citenamefont {Barends},\ and\
  \citenamefont {Martinis}}]{fowler2014scalable}%
  \BibitemOpen
  \bibfield  {author} {\bibinfo {author} {\bibfnamefont {A.~G.}\ \bibnamefont
  {Fowler}}, \bibinfo {author} {\bibfnamefont {D.}~\bibnamefont {Sank}},
  \bibinfo {author} {\bibfnamefont {J.}~\bibnamefont {Kelly}}, \bibinfo
  {author} {\bibfnamefont {R.}~\bibnamefont {Barends}},\ and\ \bibinfo {author}
  {\bibfnamefont {J.~M.}\ \bibnamefont {Martinis}},\ }\bibfield  {title}
  {\bibinfo {title} {Scalable extraction of error models from the output of
  error detection circuits},\ }\href@noop {} {\bibfield  {journal} {\bibinfo
  {journal} {arXiv preprint arXiv:1405.1454}\ } (\bibinfo {year}
  {2014})}\BibitemShut {NoStop}%
\bibitem [{\citenamefont {Takita}\ \emph {et~al.}(2016)\citenamefont {Takita},
  \citenamefont {C{\'o}rcoles}, \citenamefont {Magesan}, \citenamefont {Abdo},
  \citenamefont {Brink}, \citenamefont {Cross}, \citenamefont {Chow},\ and\
  \citenamefont {Gambetta}}]{takita2016demonstration}%
  \BibitemOpen
  \bibfield  {author} {\bibinfo {author} {\bibfnamefont {M.}~\bibnamefont
  {Takita}}, \bibinfo {author} {\bibfnamefont {A.~D.}\ \bibnamefont
  {C{\'o}rcoles}}, \bibinfo {author} {\bibfnamefont {E.}~\bibnamefont
  {Magesan}}, \bibinfo {author} {\bibfnamefont {B.}~\bibnamefont {Abdo}},
  \bibinfo {author} {\bibfnamefont {M.}~\bibnamefont {Brink}}, \bibinfo
  {author} {\bibfnamefont {A.}~\bibnamefont {Cross}}, \bibinfo {author}
  {\bibfnamefont {J.~M.}\ \bibnamefont {Chow}},\ and\ \bibinfo {author}
  {\bibfnamefont {J.~M.}\ \bibnamefont {Gambetta}},\ }\bibfield  {title}
  {\bibinfo {title} {Demonstration of weight-four parity measurements in the
  surface code architecture},\ }\href@noop {} {\bibfield  {journal} {\bibinfo
  {journal} {Physical review letters}\ }\textbf {\bibinfo {volume} {117}},\
  \bibinfo {pages} {210505} (\bibinfo {year} {2016})}\BibitemShut {NoStop}%
\bibitem [{\citenamefont {Willsch}\ \emph {et~al.}(2018)\citenamefont
  {Willsch}, \citenamefont {Willsch}, \citenamefont {Jin}, \citenamefont
  {De~Raedt},\ and\ \citenamefont {Michielsen}}]{willsch2018testing}%
  \BibitemOpen
  \bibfield  {author} {\bibinfo {author} {\bibfnamefont {D.}~\bibnamefont
  {Willsch}}, \bibinfo {author} {\bibfnamefont {M.}~\bibnamefont {Willsch}},
  \bibinfo {author} {\bibfnamefont {F.}~\bibnamefont {Jin}}, \bibinfo {author}
  {\bibfnamefont {H.}~\bibnamefont {De~Raedt}},\ and\ \bibinfo {author}
  {\bibfnamefont {K.}~\bibnamefont {Michielsen}},\ }\bibfield  {title}
  {\bibinfo {title} {Testing quantum fault tolerance on small systems},\
  }\href@noop {} {\bibfield  {journal} {\bibinfo  {journal} {Physical Review
  A}\ }\textbf {\bibinfo {volume} {98}},\ \bibinfo {pages} {052348} (\bibinfo
  {year} {2018})}\BibitemShut {NoStop}%
\bibitem [{\citenamefont {Combes}\ \emph {et~al.}(2017)\citenamefont {Combes},
  \citenamefont {Granade}, \citenamefont {Ferrie},\ and\ \citenamefont
  {Flammia}}]{Combes2017-ns}%
  \BibitemOpen
  \bibfield  {author} {\bibinfo {author} {\bibfnamefont {J.}~\bibnamefont
  {Combes}}, \bibinfo {author} {\bibfnamefont {C.}~\bibnamefont {Granade}},
  \bibinfo {author} {\bibfnamefont {C.}~\bibnamefont {Ferrie}},\ and\ \bibinfo
  {author} {\bibfnamefont {S.~T.}\ \bibnamefont {Flammia}},\ }\bibfield
  {title} {\bibinfo {title} {Logical randomized benchmarking},\ }\href
  {http://arxiv.org/abs/1702.03688} {\bibfield  {journal} {\bibinfo  {journal}
  {arxiv}\ } (\bibinfo {year} {2017})},\ \Eprint
  {https://arxiv.org/abs/1702.03688} {arXiv:1702.03688 [quant-ph]} \BibitemShut
  {NoStop}%
\bibitem [{\citenamefont {Luis}\ and\ \citenamefont
  {S{\'a}nchez-Soto}(1999)}]{Luis1999-lw}%
  \BibitemOpen
  \bibfield  {author} {\bibinfo {author} {\bibfnamefont {A.}~\bibnamefont
  {Luis}}\ and\ \bibinfo {author} {\bibfnamefont {L.~L.}\ \bibnamefont
  {S{\'a}nchez-Soto}},\ }\bibfield  {title} {\bibinfo {title} {Complete
  characterization of arbitrary quantum measurement processes},\ }\href
  {https://doi.org/10.1103/PhysRevLett.83.3573} {\bibfield  {journal} {\bibinfo
   {journal} {Phys. Rev. Lett.}\ }\textbf {\bibinfo {volume} {83}},\ \bibinfo
  {pages} {3573} (\bibinfo {year} {1999})}\BibitemShut {NoStop}%
\bibitem [{\citenamefont {Lundeen}\ \emph {et~al.}(2009)\citenamefont
  {Lundeen}, \citenamefont {Feito}, \citenamefont {Coldenstrodt-Ronge},
  \citenamefont {Pregnell}, \citenamefont {Silberhorn}, \citenamefont {Ralph},
  \citenamefont {Eisert}, \citenamefont {Plenio},\ and\ \citenamefont
  {Walmsley}}]{Lundeen2009-jo}%
  \BibitemOpen
  \bibfield  {author} {\bibinfo {author} {\bibfnamefont {J.~S.}\ \bibnamefont
  {Lundeen}}, \bibinfo {author} {\bibfnamefont {A.}~\bibnamefont {Feito}},
  \bibinfo {author} {\bibfnamefont {H.}~\bibnamefont {Coldenstrodt-Ronge}},
  \bibinfo {author} {\bibfnamefont {K.~L.}\ \bibnamefont {Pregnell}}, \bibinfo
  {author} {\bibfnamefont {C.}~\bibnamefont {Silberhorn}}, \bibinfo {author}
  {\bibfnamefont {T.~C.}\ \bibnamefont {Ralph}}, \bibinfo {author}
  {\bibfnamefont {J.}~\bibnamefont {Eisert}}, \bibinfo {author} {\bibfnamefont
  {M.~B.}\ \bibnamefont {Plenio}},\ and\ \bibinfo {author} {\bibfnamefont
  {I.~A.}\ \bibnamefont {Walmsley}},\ }\bibfield  {title} {\bibinfo {title}
  {Tomography of quantum detectors},\ }\href
  {https://doi.org/10.1038/nphys1133} {\bibfield  {journal} {\bibinfo
  {journal} {Nat. Phys.}\ }\textbf {\bibinfo {volume} {5}},\ \bibinfo {pages}
  {27} (\bibinfo {year} {2009})}\BibitemShut {NoStop}%
\bibitem [{\citenamefont {Fiur{\'a}{\v s}ek}(2001)}]{Fiurasek2001-yl}%
  \BibitemOpen
  \bibfield  {author} {\bibinfo {author} {\bibfnamefont {J.}~\bibnamefont
  {Fiur{\'a}{\v s}ek}},\ }\bibfield  {title} {\bibinfo {title}
  {Maximum-likelihood estimation of quantum measurement},\ }\href
  {https://doi.org/10.1103/PhysRevA.64.024102} {\bibfield  {journal} {\bibinfo
  {journal} {Phys. Rev. A}\ }\textbf {\bibinfo {volume} {64}},\ \bibinfo
  {pages} {024102} (\bibinfo {year} {2001})}\BibitemShut {NoStop}%
\bibitem [{\citenamefont {D'Ariano}\ \emph {et~al.}(2004)\citenamefont
  {D'Ariano}, \citenamefont {Maccone},\ and\ \citenamefont
  {Lo~Presti}}]{DAriano2004-rb}%
  \BibitemOpen
  \bibfield  {author} {\bibinfo {author} {\bibfnamefont {G.~M.}\ \bibnamefont
  {D'Ariano}}, \bibinfo {author} {\bibfnamefont {L.}~\bibnamefont {Maccone}},\
  and\ \bibinfo {author} {\bibfnamefont {P.}~\bibnamefont {Lo~Presti}},\
  }\bibfield  {title} {{\selectlanguage {english}\bibinfo {title} {Quantum
  calibration of measurement instrumentation}},\ }\href
  {https://doi.org/10.1103/PhysRevLett.93.250407} {\bibfield  {journal}
  {\bibinfo  {journal} {Phys. Rev. Lett.}\ }\textbf {\bibinfo {volume} {93}},\
  \bibinfo {pages} {250407} (\bibinfo {year} {2004})}\BibitemShut {NoStop}%
\bibitem [{\citenamefont {Izumi}\ \emph {et~al.}(2020)\citenamefont {Izumi},
  \citenamefont {Neergaard-Nielsen},\ and\ \citenamefont
  {Andersen}}]{Izumi2020-db}%
  \BibitemOpen
  \bibfield  {author} {\bibinfo {author} {\bibfnamefont {S.}~\bibnamefont
  {Izumi}}, \bibinfo {author} {\bibfnamefont {J.~S.}\ \bibnamefont
  {Neergaard-Nielsen}},\ and\ \bibinfo {author} {\bibfnamefont {U.~L.}\
  \bibnamefont {Andersen}},\ }\bibfield  {title} {\bibinfo {title} {Tomography
  of a feedback measurement with photon detection},\ }\href
  {https://doi.org/10.1103/PhysRevLett.124.070502} {\bibfield  {journal}
  {\bibinfo  {journal} {Phys. Rev. Lett.}\ }\textbf {\bibinfo {volume} {124}},\
  \bibinfo {pages} {070502} (\bibinfo {year} {2020})}\BibitemShut {NoStop}%
\bibitem [{\citenamefont {Merkel}\ \emph {et~al.}(2013)\citenamefont {Merkel},
  \citenamefont {Gambetta}, \citenamefont {Smolin}, \citenamefont {Poletto},
  \citenamefont {C{\'o}rcoles}, \citenamefont {Johnson}, \citenamefont {Ryan},\
  and\ \citenamefont {Steffen}}]{merkel2013self}%
  \BibitemOpen
  \bibfield  {author} {\bibinfo {author} {\bibfnamefont {S.~T.}\ \bibnamefont
  {Merkel}}, \bibinfo {author} {\bibfnamefont {J.~M.}\ \bibnamefont
  {Gambetta}}, \bibinfo {author} {\bibfnamefont {J.~A.}\ \bibnamefont
  {Smolin}}, \bibinfo {author} {\bibfnamefont {S.}~\bibnamefont {Poletto}},
  \bibinfo {author} {\bibfnamefont {A.~D.}\ \bibnamefont {C{\'o}rcoles}},
  \bibinfo {author} {\bibfnamefont {B.~R.}\ \bibnamefont {Johnson}}, \bibinfo
  {author} {\bibfnamefont {C.~A.}\ \bibnamefont {Ryan}},\ and\ \bibinfo
  {author} {\bibfnamefont {M.}~\bibnamefont {Steffen}},\ }\bibfield  {title}
  {\bibinfo {title} {Self-consistent quantum process tomography},\ }\href
  {https://journals.aps.org/pra/abstract/10.1103/PhysRevA.87.062119} {\bibfield
   {journal} {\bibinfo  {journal} {Phys. Rev. A}\ }\textbf {\bibinfo {volume}
  {87}},\ \bibinfo {pages} {062119} (\bibinfo {year} {2013})}\BibitemShut
  {NoStop}%
\bibitem [{\citenamefont {Davies}\ and\ \citenamefont
  {Lewis}(1970)}]{Davies1970-js}%
  \BibitemOpen
  \bibfield  {author} {\bibinfo {author} {\bibfnamefont {E.~B.}\ \bibnamefont
  {Davies}}\ and\ \bibinfo {author} {\bibfnamefont {J.~T.}\ \bibnamefont
  {Lewis}},\ }\bibfield  {title} {\bibinfo {title} {An operational approach to
  quantum probability},\ }\href {https://doi.org/10.1007/BF01647093} {\bibfield
   {journal} {\bibinfo  {journal} {Commun. Math. Phys.}\ }\textbf {\bibinfo
  {volume} {17}},\ \bibinfo {pages} {239} (\bibinfo {year} {1970})}\BibitemShut
  {NoStop}%
\bibitem [{\citenamefont {Miklin}\ \emph {et~al.}(2020)\citenamefont {Miklin},
  \citenamefont {Borka{\l}a},\ and\ \citenamefont
  {Paw{\l}owski}}]{Miklin2020-pn}%
  \BibitemOpen
  \bibfield  {author} {\bibinfo {author} {\bibfnamefont {N.}~\bibnamefont
  {Miklin}}, \bibinfo {author} {\bibfnamefont {J.~J.}\ \bibnamefont
  {Borka{\l}a}},\ and\ \bibinfo {author} {\bibfnamefont {M.}~\bibnamefont
  {Paw{\l}owski}},\ }\bibfield  {title} {\bibinfo {title}
  {Semi-device-independent self-testing of unsharp measurements},\ }\href
  {https://doi.org/10.1103/PhysRevResearch.2.033014} {\bibfield  {journal}
  {\bibinfo  {journal} {Phys. Rev. Research}\ }\textbf {\bibinfo {volume}
  {2}},\ \bibinfo {pages} {033014} (\bibinfo {year} {2020})}\BibitemShut
  {NoStop}%
\bibitem [{\citenamefont {Mohan}\ \emph {et~al.}(2019)\citenamefont {Mohan},
  \citenamefont {Tavakoli},\ and\ \citenamefont {Brunner}}]{Mohan2019-xe}%
  \BibitemOpen
  \bibfield  {author} {\bibinfo {author} {\bibfnamefont {K.}~\bibnamefont
  {Mohan}}, \bibinfo {author} {\bibfnamefont {A.}~\bibnamefont {Tavakoli}},\
  and\ \bibinfo {author} {\bibfnamefont {N.}~\bibnamefont {Brunner}},\
  }\bibfield  {title} {\bibinfo {title} {Sequential random access codes and
  self-testing of quantum measurement instruments},\ }\href
  {https://doi.org/10.1088/1367-2630/ab3773} {\bibfield  {journal} {\bibinfo
  {journal} {New J. Phys.}\ }\textbf {\bibinfo {volume} {21}},\ \bibinfo
  {pages} {083034} (\bibinfo {year} {2019})}\BibitemShut {NoStop}%
\bibitem [{\citenamefont {Wagner}\ \emph {et~al.}(2020)\citenamefont {Wagner},
  \citenamefont {Bancal}, \citenamefont {Sangouard},\ and\ \citenamefont
  {Sekatski}}]{Wagner2020-hl}%
  \BibitemOpen
  \bibfield  {author} {\bibinfo {author} {\bibfnamefont {S.}~\bibnamefont
  {Wagner}}, \bibinfo {author} {\bibfnamefont {J.-D.}\ \bibnamefont {Bancal}},
  \bibinfo {author} {\bibfnamefont {N.}~\bibnamefont {Sangouard}},\ and\
  \bibinfo {author} {\bibfnamefont {P.}~\bibnamefont {Sekatski}},\ }\bibfield
  {title} {\bibinfo {title} {Device-independent characterization of quantum
  instruments},\ }\href {https://doi.org/10.22331/q-2020-03-19-243} {\bibfield
  {journal} {\bibinfo  {journal} {Quantum}\ }\textbf {\bibinfo {volume} {4}},\
  \bibinfo {pages} {243} (\bibinfo {year} {2020})}\BibitemShut {NoStop}%
\bibitem [{\citenamefont {Blumoff}\ \emph {et~al.}(2016)\citenamefont
  {Blumoff}, \citenamefont {Chou}, \citenamefont {Shen}, \citenamefont
  {Reagor}, \citenamefont {Axline}, \citenamefont {Brierley}, \citenamefont
  {Silveri}, \citenamefont {Wang}, \citenamefont {Vlastakis}, \citenamefont
  {Nigg}, \citenamefont {Frunzio}, \citenamefont {Devoret}, \citenamefont
  {Jiang}, \citenamefont {Girvin},\ and\ \citenamefont
  {Schoelkopf}}]{Blumoff2016-gy}%
  \BibitemOpen
  \bibfield  {author} {\bibinfo {author} {\bibfnamefont {J.~Z.}\ \bibnamefont
  {Blumoff}}, \bibinfo {author} {\bibfnamefont {K.}~\bibnamefont {Chou}},
  \bibinfo {author} {\bibfnamefont {C.}~\bibnamefont {Shen}}, \bibinfo {author}
  {\bibfnamefont {M.}~\bibnamefont {Reagor}}, \bibinfo {author} {\bibfnamefont
  {C.}~\bibnamefont {Axline}}, \bibinfo {author} {\bibfnamefont {R.~T.}\
  \bibnamefont {Brierley}}, \bibinfo {author} {\bibfnamefont {M.~P.}\
  \bibnamefont {Silveri}}, \bibinfo {author} {\bibfnamefont {C.}~\bibnamefont
  {Wang}}, \bibinfo {author} {\bibfnamefont {B.}~\bibnamefont {Vlastakis}},
  \bibinfo {author} {\bibfnamefont {S.~E.}\ \bibnamefont {Nigg}}, \bibinfo
  {author} {\bibfnamefont {L.}~\bibnamefont {Frunzio}}, \bibinfo {author}
  {\bibfnamefont {M.~H.}\ \bibnamefont {Devoret}}, \bibinfo {author}
  {\bibfnamefont {L.}~\bibnamefont {Jiang}}, \bibinfo {author} {\bibfnamefont
  {S.~M.}\ \bibnamefont {Girvin}},\ and\ \bibinfo {author} {\bibfnamefont
  {R.~J.}\ \bibnamefont {Schoelkopf}},\ }\bibfield  {title} {\bibinfo {title}
  {Implementing and characterizing precise multiqubit measurements},\ }\href
  {https://doi.org/10.1103/PhysRevX.6.031041} {\bibfield  {journal} {\bibinfo
  {journal} {Phys. Rev. X}\ }\textbf {\bibinfo {volume} {6}},\ \bibinfo {pages}
  {031041} (\bibinfo {year} {2016})}\BibitemShut {NoStop}%
\bibitem [{\citenamefont {Horodecki}\ \emph {et~al.}(2003)\citenamefont
  {Horodecki}, \citenamefont {Shor},\ and\ \citenamefont
  {Ruskai}}]{Horodecki2003-fm}%
  \BibitemOpen
  \bibfield  {author} {\bibinfo {author} {\bibfnamefont {M.}~\bibnamefont
  {Horodecki}}, \bibinfo {author} {\bibfnamefont {P.~W.}\ \bibnamefont
  {Shor}},\ and\ \bibinfo {author} {\bibfnamefont {M.~B.}\ \bibnamefont
  {Ruskai}},\ }\bibfield  {title} {\bibinfo {title} {Entanglement breaking
  channels},\ }\href {https://doi.org/10.1142/S0129055X03001709} {\bibfield
  {journal} {\bibinfo  {journal} {Rev. Math. Phys.}\ }\textbf {\bibinfo
  {volume} {15}},\ \bibinfo {pages} {629} (\bibinfo {year} {2003})}\BibitemShut
  {NoStop}%
\bibitem [{sup()}]{supplement}%
  \BibitemOpen
  \href@noop {} {\bibinfo {title} {See supplemental material.}}\BibitemShut
  {Stop}%
\bibitem [{\citenamefont {Aharonov}\ \emph {et~al.}(1998)\citenamefont
  {Aharonov}, \citenamefont {Kitaev},\ and\ \citenamefont
  {Nisan}}]{Aharonov1998-ov}%
  \BibitemOpen
  \bibfield  {author} {\bibinfo {author} {\bibfnamefont {D.}~\bibnamefont
  {Aharonov}}, \bibinfo {author} {\bibfnamefont {A.}~\bibnamefont {Kitaev}},\
  and\ \bibinfo {author} {\bibfnamefont {N.}~\bibnamefont {Nisan}},\ }\bibfield
   {title} {\bibinfo {title} {Quantum circuits with mixed states},\ }in\ \href
  {https://doi.org/10.1145/276698.276708} {\emph {\bibinfo {booktitle}
  {Proceedings of the thirtieth annual {ACM} symposium on Theory of computing -
  {STOC} '98}}}\ (\bibinfo  {publisher} {ACM Press},\ \bibinfo {address} {New
  York, New York, USA},\ \bibinfo {year} {1998})\ pp.\ \bibinfo {pages}
  {20--30}\BibitemShut {NoStop}%
\bibitem [{\citenamefont {O'Brien}\ \emph {et~al.}(2004)\citenamefont
  {O'Brien}, \citenamefont {Pryde}, \citenamefont {Gilchrist}, \citenamefont
  {James},\ and\ \citenamefont {{others}}}]{OBrien2004-tr}%
  \BibitemOpen
  \bibfield  {author} {\bibinfo {author} {\bibfnamefont {J.~L.}\ \bibnamefont
  {O'Brien}}, \bibinfo {author} {\bibfnamefont {G.~J.}\ \bibnamefont {Pryde}},
  \bibinfo {author} {\bibfnamefont {A.}~\bibnamefont {Gilchrist}}, \bibinfo
  {author} {\bibfnamefont {D.~F.~V.}\ \bibnamefont {James}},\ and\ \bibinfo
  {author} {\bibnamefont {{others}}},\ }\bibfield  {title} {\bibinfo {title}
  {Quantum process tomography of a {controlled-NOT} gate},\ }\href@noop {}
  {\bibfield  {journal} {\bibinfo  {journal} {Physical review}\ } (\bibinfo
  {year} {2004})}\BibitemShut {NoStop}%
\bibitem [{\citenamefont {Poyatos}\ \emph {et~al.}(1997)\citenamefont
  {Poyatos}, \citenamefont {Cirac},\ and\ \citenamefont
  {Zoller}}]{Poyatos1997-mz}%
  \BibitemOpen
  \bibfield  {author} {\bibinfo {author} {\bibfnamefont {J.~F.}\ \bibnamefont
  {Poyatos}}, \bibinfo {author} {\bibfnamefont {J.~I.}\ \bibnamefont {Cirac}},\
  and\ \bibinfo {author} {\bibfnamefont {P.}~\bibnamefont {Zoller}},\
  }\bibfield  {title} {\bibinfo {title} {Complete characterization of a quantum
  process: The {Two-Bit} quantum gate},\ }\href
  {https://doi.org/10.1103/PhysRevLett.78.390} {\bibfield  {journal} {\bibinfo
  {journal} {Phys. Rev. Lett.}\ }\textbf {\bibinfo {volume} {78}},\ \bibinfo
  {pages} {390} (\bibinfo {year} {1997})}\BibitemShut {NoStop}%
\bibitem [{\citenamefont {Chuang}\ and\ \citenamefont
  {Nielsen}(1997)}]{Chuang1997-vf}%
  \BibitemOpen
  \bibfield  {author} {\bibinfo {author} {\bibfnamefont {I.~L.}\ \bibnamefont
  {Chuang}}\ and\ \bibinfo {author} {\bibfnamefont {M.~A.}\ \bibnamefont
  {Nielsen}},\ }\bibfield  {title} {\bibinfo {title} {Prescription for
  experimental determination of the dynamics of a quantum black box},\ }\href
  {https://doi.org/10.1080/09500349708231894} {\bibfield  {journal} {\bibinfo
  {journal} {J. Mod. Opt.}\ }\textbf {\bibinfo {volume} {44}},\ \bibinfo
  {pages} {2455} (\bibinfo {year} {1997})}\BibitemShut {NoStop}%
\bibitem [{\citenamefont {Nielsen}\ \emph {et~al.}(2019)\citenamefont
  {Nielsen}, \citenamefont {Blume-Kohout}, \citenamefont {Rudinger},
  \citenamefont {Proctor}, \citenamefont {Saldyt},\ and\ \citenamefont
  {{Others}}}]{Nielsen2019-wi}%
  \BibitemOpen
  \bibfield  {author} {\bibinfo {author} {\bibfnamefont {E.}~\bibnamefont
  {Nielsen}}, \bibinfo {author} {\bibfnamefont {R.~J.}\ \bibnamefont
  {Blume-Kohout}}, \bibinfo {author} {\bibfnamefont {K.~M.}\ \bibnamefont
  {Rudinger}}, \bibinfo {author} {\bibfnamefont {T.~J.}\ \bibnamefont
  {Proctor}}, \bibinfo {author} {\bibfnamefont {L.}~\bibnamefont {Saldyt}},\
  and\ \bibinfo {author} {\bibnamefont {{Others}}},\ }\href@noop {} {\emph
  {\bibinfo {title} {Python {GST} Implementation ({PyGSTi}) v. 0.9}}},\
  \bibinfo {type} {Tech. Rep.}\ (\bibinfo  {institution} {Sandia National
  Lab.(SNL-NM), Albuquerque, NM (United States)},\ \bibinfo {year}
  {2019})\BibitemShut {NoStop}%
\bibitem [{\citenamefont {Magesan}\ \emph {et~al.}(2011)\citenamefont
  {Magesan}, \citenamefont {Gambetta},\ and\ \citenamefont
  {Emerson}}]{Magesan2011-hc}%
  \BibitemOpen
  \bibfield  {author} {\bibinfo {author} {\bibfnamefont {E.}~\bibnamefont
  {Magesan}}, \bibinfo {author} {\bibfnamefont {J.~M.}\ \bibnamefont
  {Gambetta}},\ and\ \bibinfo {author} {\bibfnamefont {J.}~\bibnamefont
  {Emerson}},\ }\bibfield  {title} {{\selectlanguage {english}\bibinfo {title}
  {Scalable and robust randomized benchmarking of quantum processes}},\ }\href
  {https://doi.org/10.1103/PhysRevLett.106.180504} {\bibfield  {journal}
  {\bibinfo  {journal} {Phys. Rev. Lett.}\ }\textbf {\bibinfo {volume} {106}},\
  \bibinfo {pages} {180504} (\bibinfo {year} {2011})}\BibitemShut {NoStop}%
\bibitem [{\citenamefont {Blais}\ \emph {et~al.}(2004)\citenamefont {Blais},
  \citenamefont {Huang}, \citenamefont {Wallraff}, \citenamefont {Girvin},\
  and\ \citenamefont {Schoelkopf}}]{Blais:2004}%
  \BibitemOpen
  \bibfield  {author} {\bibinfo {author} {\bibfnamefont {A.}~\bibnamefont
  {Blais}}, \bibinfo {author} {\bibfnamefont {R.-S.}\ \bibnamefont {Huang}},
  \bibinfo {author} {\bibfnamefont {A.}~\bibnamefont {Wallraff}}, \bibinfo
  {author} {\bibfnamefont {S.~M.}\ \bibnamefont {Girvin}},\ and\ \bibinfo
  {author} {\bibfnamefont {R.~J.}\ \bibnamefont {Schoelkopf}},\ }\bibfield
  {title} {\bibinfo {title} {Cavity quantum electrodynamics for superconducting
  electrical circuits: An architecture for quantum computation},\ }\href
  {https://doi.org/10.1103/PhysRevA.69.062320} {\bibfield  {journal} {\bibinfo
  {journal} {Phys. Rev. A}\ }\textbf {\bibinfo {volume} {69}},\ \bibinfo
  {pages} {062320} (\bibinfo {year} {2004})}\BibitemShut {NoStop}%
\bibitem [{\citenamefont {O'Brien}\ \emph {et~al.}(2014)\citenamefont
  {O'Brien}, \citenamefont {Macklin}, \citenamefont {Siddiqi},\ and\
  \citenamefont {Zhang}}]{OBrian:2014}%
  \BibitemOpen
  \bibfield  {author} {\bibinfo {author} {\bibfnamefont {K.}~\bibnamefont
  {O'Brien}}, \bibinfo {author} {\bibfnamefont {C.}~\bibnamefont {Macklin}},
  \bibinfo {author} {\bibfnamefont {I.}~\bibnamefont {Siddiqi}},\ and\ \bibinfo
  {author} {\bibfnamefont {X.}~\bibnamefont {Zhang}},\ }\bibfield  {title}
  {\bibinfo {title} {Resonant phase matching of josephson junction traveling
  wave parametric amplifiers},\ }\href
  {https://doi.org/10.1103/PhysRevLett.113.157001} {\bibfield  {journal}
  {\bibinfo  {journal} {Phys. Rev. Lett.}\ }\textbf {\bibinfo {volume} {113}},\
  \bibinfo {pages} {157001} (\bibinfo {year} {2014})}\BibitemShut {NoStop}%
\bibitem [{\citenamefont {Ryan}\ \emph {et~al.}(2015)\citenamefont {Ryan},
  \citenamefont {Johnson}, \citenamefont {Gambetta}, \citenamefont {Chow},
  \citenamefont {da~Silva}, \citenamefont {Dial},\ and\ \citenamefont
  {Ohki}}]{ryan2015tomography}%
  \BibitemOpen
  \bibfield  {author} {\bibinfo {author} {\bibfnamefont {C.~A.}\ \bibnamefont
  {Ryan}}, \bibinfo {author} {\bibfnamefont {B.~R.}\ \bibnamefont {Johnson}},
  \bibinfo {author} {\bibfnamefont {J.~M.}\ \bibnamefont {Gambetta}}, \bibinfo
  {author} {\bibfnamefont {J.~M.}\ \bibnamefont {Chow}}, \bibinfo {author}
  {\bibfnamefont {M.~P.}\ \bibnamefont {da~Silva}}, \bibinfo {author}
  {\bibfnamefont {O.~E.}\ \bibnamefont {Dial}},\ and\ \bibinfo {author}
  {\bibfnamefont {T.~A.}\ \bibnamefont {Ohki}},\ }\bibfield  {title} {\bibinfo
  {title} {Tomography via correlation of noisy measurement records},\
  }\href@noop {} {\bibfield  {journal} {\bibinfo  {journal} {Physical Review
  A}\ }\textbf {\bibinfo {volume} {91}},\ \bibinfo {pages} {022118} (\bibinfo
  {year} {2015})}\BibitemShut {NoStop}%
\bibitem [{\citenamefont {McClure}\ \emph {et~al.}(2016)\citenamefont
  {McClure}, \citenamefont {Paik}, \citenamefont {Bishop}, \citenamefont
  {Steffen}, \citenamefont {Chow},\ and\ \citenamefont
  {Gambetta}}]{McClure2016}%
  \BibitemOpen
  \bibfield  {author} {\bibinfo {author} {\bibfnamefont {D.~T.}\ \bibnamefont
  {McClure}}, \bibinfo {author} {\bibfnamefont {H.}~\bibnamefont {Paik}},
  \bibinfo {author} {\bibfnamefont {L.~S.}\ \bibnamefont {Bishop}}, \bibinfo
  {author} {\bibfnamefont {M.}~\bibnamefont {Steffen}}, \bibinfo {author}
  {\bibfnamefont {J.~M.}\ \bibnamefont {Chow}},\ and\ \bibinfo {author}
  {\bibfnamefont {J.~M.}\ \bibnamefont {Gambetta}},\ }\bibfield  {title}
  {\bibinfo {title} {{Rapid Driven Reset of a Qubit Readout Resonator}},\
  }\href {https://doi.org/10.1103/PhysRevApplied.5.011001} {\bibfield
  {journal} {\bibinfo  {journal} {Phys. Rev. Appl.}\ }\textbf {\bibinfo
  {volume} {5}},\ \bibinfo {pages} {011001} (\bibinfo {year} {2016})},\ \Eprint
  {https://arxiv.org/abs/1503.01456} {arXiv:1503.01456} \BibitemShut {NoStop}%
\bibitem [{\citenamefont {Wan}\ \emph {et~al.}(2019)\citenamefont {Wan},
  \citenamefont {Kienzler}, \citenamefont {Erickson}, \citenamefont {Mayer},
  \citenamefont {Tan}, \citenamefont {Wu}, \citenamefont {Vasconcelos},
  \citenamefont {Glancy}, \citenamefont {Knill}, \citenamefont {Wineland},
  \citenamefont {Wilson},\ and\ \citenamefont {Leibfried}}]{Wan2019-rn}%
  \BibitemOpen
  \bibfield  {author} {\bibinfo {author} {\bibfnamefont {Y.}~\bibnamefont
  {Wan}}, \bibinfo {author} {\bibfnamefont {D.}~\bibnamefont {Kienzler}},
  \bibinfo {author} {\bibfnamefont {S.~D.}\ \bibnamefont {Erickson}}, \bibinfo
  {author} {\bibfnamefont {K.~H.}\ \bibnamefont {Mayer}}, \bibinfo {author}
  {\bibfnamefont {T.~R.}\ \bibnamefont {Tan}}, \bibinfo {author} {\bibfnamefont
  {J.~J.}\ \bibnamefont {Wu}}, \bibinfo {author} {\bibfnamefont {H.~M.}\
  \bibnamefont {Vasconcelos}}, \bibinfo {author} {\bibfnamefont
  {S.}~\bibnamefont {Glancy}}, \bibinfo {author} {\bibfnamefont
  {E.}~\bibnamefont {Knill}}, \bibinfo {author} {\bibfnamefont {D.~J.}\
  \bibnamefont {Wineland}}, \bibinfo {author} {\bibfnamefont {A.~C.}\
  \bibnamefont {Wilson}},\ and\ \bibinfo {author} {\bibfnamefont
  {D.}~\bibnamefont {Leibfried}},\ }\bibfield  {title} {\bibinfo {title}
  {Quantum gate teleportation between separated qubits in a trapped-ion
  processor},\ }\href {https://doi.org/10.1126/science.aaw9415} {\bibfield
  {journal} {\bibinfo  {journal} {Science}\ }\textbf {\bibinfo {volume}
  {364}},\ \bibinfo {pages} {875} (\bibinfo {year} {2019})}\BibitemShut
  {NoStop}%
\bibitem [{\citenamefont {Blume-Kohout}\ \emph {et~al.}(2020)\citenamefont
  {Blume-Kohout}, \citenamefont {Rudinger}, \citenamefont {Nielsen},
  \citenamefont {Proctor},\ and\ \citenamefont {Young}}]{Blume-Kohout2020-fl}%
  \BibitemOpen
  \bibfield  {author} {\bibinfo {author} {\bibfnamefont {R.}~\bibnamefont
  {Blume-Kohout}}, \bibinfo {author} {\bibfnamefont {K.}~\bibnamefont
  {Rudinger}}, \bibinfo {author} {\bibfnamefont {E.}~\bibnamefont {Nielsen}},
  \bibinfo {author} {\bibfnamefont {T.}~\bibnamefont {Proctor}},\ and\ \bibinfo
  {author} {\bibfnamefont {K.}~\bibnamefont {Young}},\ }\bibfield  {title}
  {\bibinfo {title} {Wildcard error: Quantifying unmodeled errors in quantum
  processors},\ }\href {http://arxiv.org/abs/2012.12231} {\bibfield  {journal}
  {\bibinfo  {journal} {arxiv}\ } (\bibinfo {year} {2020})},\ \Eprint
  {https://arxiv.org/abs/2012.12231} {arXiv:2012.12231 [quant-ph]} \BibitemShut
  {NoStop}%
\bibitem [{\citenamefont {Rudinger}\ \emph {et~al.}(2019)\citenamefont
  {Rudinger}, \citenamefont {Proctor}, \citenamefont {Langharst}, \citenamefont
  {Sarovar}, \citenamefont {Young},\ and\ \citenamefont
  {Blume-Kohout}}]{Rudinger2019-ua}%
  \BibitemOpen
  \bibfield  {author} {\bibinfo {author} {\bibfnamefont {K.}~\bibnamefont
  {Rudinger}}, \bibinfo {author} {\bibfnamefont {T.}~\bibnamefont {Proctor}},
  \bibinfo {author} {\bibfnamefont {D.}~\bibnamefont {Langharst}}, \bibinfo
  {author} {\bibfnamefont {M.}~\bibnamefont {Sarovar}}, \bibinfo {author}
  {\bibfnamefont {K.}~\bibnamefont {Young}},\ and\ \bibinfo {author}
  {\bibfnamefont {R.}~\bibnamefont {Blume-Kohout}},\ }\bibfield  {title}
  {\bibinfo {title} {Probing context-dependent errors in quantum processors},\
  }\href {https://doi.org/10.1103/PhysRevX.9.021045} {\bibfield  {journal}
  {\bibinfo  {journal} {Phys. Rev. X}\ }\textbf {\bibinfo {volume} {9}},\
  \bibinfo {pages} {021045} (\bibinfo {year} {2019})}\BibitemShut {NoStop}%
\bibitem [{\citenamefont {Proctor}\ \emph {et~al.}(2020)\citenamefont
  {Proctor}, \citenamefont {Revelle}, \citenamefont {Nielsen}, \citenamefont
  {Rudinger}, \citenamefont {Lobser}, \citenamefont {Maunz}, \citenamefont
  {Blume-Kohout},\ and\ \citenamefont {Young}}]{Proctor2020-iz}%
  \BibitemOpen
  \bibfield  {author} {\bibinfo {author} {\bibfnamefont {T.}~\bibnamefont
  {Proctor}}, \bibinfo {author} {\bibfnamefont {M.}~\bibnamefont {Revelle}},
  \bibinfo {author} {\bibfnamefont {E.}~\bibnamefont {Nielsen}}, \bibinfo
  {author} {\bibfnamefont {K.}~\bibnamefont {Rudinger}}, \bibinfo {author}
  {\bibfnamefont {D.}~\bibnamefont {Lobser}}, \bibinfo {author} {\bibfnamefont
  {P.}~\bibnamefont {Maunz}}, \bibinfo {author} {\bibfnamefont
  {R.}~\bibnamefont {Blume-Kohout}},\ and\ \bibinfo {author} {\bibfnamefont
  {K.}~\bibnamefont {Young}},\ }\bibfield  {title} {\bibinfo {title} {Detecting
  and tracking drift in quantum information processors},\ }\href
  {https://doi.org/10.1038/s41467-020-19074-4} {\bibfield  {journal} {\bibinfo
  {journal} {Nat. Commun.}\ }\textbf {\bibinfo {volume} {11}},\ \bibinfo
  {pages} {5396} (\bibinfo {year} {2020})}\BibitemShut {NoStop}%
\bibitem [{\citenamefont {Bylander}\ \emph {et~al.}(2011)\citenamefont
  {Bylander}, \citenamefont {Gustavsson}, \citenamefont {Yan}, \citenamefont
  {Yoshihara}, \citenamefont {Harrabi}, \citenamefont {Fitch}, \citenamefont
  {Cory}, \citenamefont {Nakamura}, \citenamefont {Tsai},\ and\ \citenamefont
  {Oliver}}]{Bylander2011-cf}%
  \BibitemOpen
  \bibfield  {author} {\bibinfo {author} {\bibfnamefont {J.}~\bibnamefont
  {Bylander}}, \bibinfo {author} {\bibfnamefont {S.}~\bibnamefont
  {Gustavsson}}, \bibinfo {author} {\bibfnamefont {F.}~\bibnamefont {Yan}},
  \bibinfo {author} {\bibfnamefont {F.}~\bibnamefont {Yoshihara}}, \bibinfo
  {author} {\bibfnamefont {K.}~\bibnamefont {Harrabi}}, \bibinfo {author}
  {\bibfnamefont {G.}~\bibnamefont {Fitch}}, \bibinfo {author} {\bibfnamefont
  {D.~G.}\ \bibnamefont {Cory}}, \bibinfo {author} {\bibfnamefont
  {Y.}~\bibnamefont {Nakamura}}, \bibinfo {author} {\bibfnamefont {J.-S.}\
  \bibnamefont {Tsai}},\ and\ \bibinfo {author} {\bibfnamefont {W.~D.}\
  \bibnamefont {Oliver}},\ }\bibfield  {title} {\bibinfo {title} {Noise
  spectroscopy through dynamical decoupling with a superconducting flux
  qubit},\ }\href {https://doi.org/10.1038/nphys1994} {\bibfield  {journal}
  {\bibinfo  {journal} {Nat. Phys.}\ }\textbf {\bibinfo {volume} {7}},\
  \bibinfo {pages} {565} (\bibinfo {year} {2011})}\BibitemShut {NoStop}%
\bibitem [{\citenamefont {Nielsen}\ \emph {et~al.}(2021)\citenamefont
  {Nielsen}, \citenamefont {Rudinger}, \citenamefont {Proctor}, \citenamefont
  {Young},\ and\ \citenamefont {Blume-Kohout}}]{nielsen2021efficient}%
  \BibitemOpen
  \bibfield  {author} {\bibinfo {author} {\bibfnamefont {E.}~\bibnamefont
  {Nielsen}}, \bibinfo {author} {\bibfnamefont {K.}~\bibnamefont {Rudinger}},
  \bibinfo {author} {\bibfnamefont {T.}~\bibnamefont {Proctor}}, \bibinfo
  {author} {\bibfnamefont {K.}~\bibnamefont {Young}},\ and\ \bibinfo {author}
  {\bibfnamefont {R.}~\bibnamefont {Blume-Kohout}},\ }\href@noop {} {\bibinfo
  {title} {Efficient flexible characterization of quantum processors with
  nested error models}} (\bibinfo {year} {2021}),\ \Eprint
  {https://arxiv.org/abs/2103.02188} {arXiv:2103.02188 [quant-ph]} \BibitemShut
  {NoStop}%
\bibitem [{\citenamefont {Bultink}\ \emph {et~al.}(2018)\citenamefont
  {Bultink}, \citenamefont {Tarasinski}, \citenamefont {Haandbæk},
  \citenamefont {Poletto}, \citenamefont {Haider}, \citenamefont {Michalak},
  \citenamefont {Bruno},\ and\ \citenamefont {DiCarlo}}]{Bultink2018}%
  \BibitemOpen
  \bibfield  {author} {\bibinfo {author} {\bibfnamefont {C.~C.}\ \bibnamefont
  {Bultink}}, \bibinfo {author} {\bibfnamefont {B.}~\bibnamefont {Tarasinski}},
  \bibinfo {author} {\bibfnamefont {N.}~\bibnamefont {Haandbæk}}, \bibinfo
  {author} {\bibfnamefont {S.}~\bibnamefont {Poletto}}, \bibinfo {author}
  {\bibfnamefont {N.}~\bibnamefont {Haider}}, \bibinfo {author} {\bibfnamefont
  {D.~J.}\ \bibnamefont {Michalak}}, \bibinfo {author} {\bibfnamefont
  {A.}~\bibnamefont {Bruno}},\ and\ \bibinfo {author} {\bibfnamefont
  {L.}~\bibnamefont {DiCarlo}},\ }\bibfield  {title} {\bibinfo {title} {General
  method for extracting the quantum efficiency of dispersive qubit readout in
  circuit qed},\ }\href {https://doi.org/10.1063/1.5015954} {\bibfield
  {journal} {\bibinfo  {journal} {Applied Physics Letters}\ }\textbf {\bibinfo
  {volume} {112}},\ \bibinfo {pages} {092601} (\bibinfo {year}
  {2018})}\BibitemShut {NoStop}%
\bibitem [{\citenamefont {Boissonneault}\ \emph {et~al.}(2008)\citenamefont
  {Boissonneault}, \citenamefont {Gambetta},\ and\ \citenamefont
  {Blais}}]{Boisson2008}%
  \BibitemOpen
  \bibfield  {author} {\bibinfo {author} {\bibfnamefont {M.}~\bibnamefont
  {Boissonneault}}, \bibinfo {author} {\bibfnamefont {J.~M.}\ \bibnamefont
  {Gambetta}},\ and\ \bibinfo {author} {\bibfnamefont {A.}~\bibnamefont
  {Blais}},\ }\bibfield  {title} {\bibinfo {title} {Nonlinear dispersive regime
  of cavity qed: The dressed dephasing model},\ }\href
  {https://doi.org/10.1103/PhysRevA.77.060305} {\bibfield  {journal} {\bibinfo
  {journal} {Phys. Rev. A}\ }\textbf {\bibinfo {volume} {77}},\ \bibinfo
  {pages} {060305} (\bibinfo {year} {2008})}\BibitemShut {NoStop}%
\bibitem [{\citenamefont {Boissonneault}\ \emph {et~al.}(2010)\citenamefont
  {Boissonneault}, \citenamefont {Gambetta},\ and\ \citenamefont
  {Blais}}]{Boisson2010}%
  \BibitemOpen
  \bibfield  {author} {\bibinfo {author} {\bibfnamefont {M.}~\bibnamefont
  {Boissonneault}}, \bibinfo {author} {\bibfnamefont {J.~M.}\ \bibnamefont
  {Gambetta}},\ and\ \bibinfo {author} {\bibfnamefont {A.}~\bibnamefont
  {Blais}},\ }\bibfield  {title} {\bibinfo {title} {Improved superconducting
  qubit readout by qubit-induced nonlinearities},\ }\href
  {https://doi.org/10.1103/PhysRevLett.105.100504} {\bibfield  {journal}
  {\bibinfo  {journal} {Phys. Rev. Lett.}\ }\textbf {\bibinfo {volume} {105}},\
  \bibinfo {pages} {100504} (\bibinfo {year} {2010})}\BibitemShut {NoStop}%
\bibitem [{\citenamefont {Govia}\ and\ \citenamefont
  {Wilhelm}(2015)}]{Govia2015}%
  \BibitemOpen
  \bibfield  {author} {\bibinfo {author} {\bibfnamefont {L.~C.~G.}\
  \bibnamefont {Govia}}\ and\ \bibinfo {author} {\bibfnamefont {F.~K.}\
  \bibnamefont {Wilhelm}},\ }\bibfield  {title} {\bibinfo {title}
  {Unitary-feedback-improved qubit initialization in the dispersive regime},\
  }\href {https://doi.org/10.1103/PhysRevApplied.4.054001} {\bibfield
  {journal} {\bibinfo  {journal} {Phys. Rev. Applied}\ }\textbf {\bibinfo
  {volume} {4}},\ \bibinfo {pages} {054001} (\bibinfo {year}
  {2015})}\BibitemShut {NoStop}%
\bibitem [{\citenamefont {Govia}\ and\ \citenamefont
  {Wilhelm}(2016)}]{Govia:2016}%
  \BibitemOpen
  \bibfield  {author} {\bibinfo {author} {\bibfnamefont {L.~C.~G.}\
  \bibnamefont {Govia}}\ and\ \bibinfo {author} {\bibfnamefont {F.~K.}\
  \bibnamefont {Wilhelm}},\ }\bibfield  {title} {\bibinfo {title} {Entanglement
  generated by the dispersive interaction: The dressed coherent state},\ }\href
  {https://doi.org/10.1103/PhysRevA.93.012316} {\bibfield  {journal} {\bibinfo
  {journal} {Phys. Rev. A}\ }\textbf {\bibinfo {volume} {93}},\ \bibinfo
  {pages} {012316} (\bibinfo {year} {2016})}\BibitemShut {NoStop}%
\bibitem [{\citenamefont {Khezri}\ \emph {et~al.}(2016)\citenamefont {Khezri},
  \citenamefont {Mlinar}, \citenamefont {Dressel},\ and\ \citenamefont
  {Korotkov}}]{Khezri2016}%
  \BibitemOpen
  \bibfield  {author} {\bibinfo {author} {\bibfnamefont {M.}~\bibnamefont
  {Khezri}}, \bibinfo {author} {\bibfnamefont {E.}~\bibnamefont {Mlinar}},
  \bibinfo {author} {\bibfnamefont {J.}~\bibnamefont {Dressel}},\ and\ \bibinfo
  {author} {\bibfnamefont {A.~N.}\ \bibnamefont {Korotkov}},\ }\bibfield
  {title} {\bibinfo {title} {Measuring a transmon qubit in circuit qed: Dressed
  squeezed states},\ }\href {https://doi.org/10.1103/PhysRevA.94.012347}
  {\bibfield  {journal} {\bibinfo  {journal} {Phys. Rev. A}\ }\textbf {\bibinfo
  {volume} {94}},\ \bibinfo {pages} {012347} (\bibinfo {year}
  {2016})}\BibitemShut {NoStop}%
\bibitem [{\citenamefont {Huembeli}\ and\ \citenamefont
  {Nigg}(2017)}]{Huembeli2017}%
  \BibitemOpen
  \bibfield  {author} {\bibinfo {author} {\bibfnamefont {P.}~\bibnamefont
  {Huembeli}}\ and\ \bibinfo {author} {\bibfnamefont {S.~E.}\ \bibnamefont
  {Nigg}},\ }\bibfield  {title} {\bibinfo {title} {Towards a heralded
  eigenstate-preserving measurement of multi-qubit parity in circuit qed},\
  }\href {https://doi.org/10.1103/PhysRevA.96.012313} {\bibfield  {journal}
  {\bibinfo  {journal} {Phys. Rev. A}\ }\textbf {\bibinfo {volume} {96}},\
  \bibinfo {pages} {012313} (\bibinfo {year} {2017})}\BibitemShut {NoStop}%
\bibitem [{\citenamefont {Nielsen}(2020)}]{Nielsen2020-co}%
  \BibitemOpen
  \bibfield  {author} {\bibinfo {author} {\bibfnamefont {E.}~\bibnamefont
  {Nielsen}},\ }\href {https://doi.org/10.2172/1673168} {\emph {\bibinfo
  {title} {Efficient Scalable Tomography of {Many-Qubit} Quantum
  Processors}}},\ \bibinfo {type} {Tech. Rep.}\ \bibinfo {number}
  {SAND2020-10518}\ (\bibinfo  {institution} {Sandia National Lab. (SNL-NM),
  Albuquerque, NM (United States)},\ \bibinfo {year} {2020})\BibitemShut
  {NoStop}%
\bibitem [{\citenamefont {Govia}\ \emph {et~al.}(2020)\citenamefont {Govia},
  \citenamefont {Ribeill}, \citenamefont {Rist{\`e}}, \citenamefont {Ware},\
  and\ \citenamefont {Krovi}}]{govia2020bootstrapping}%
  \BibitemOpen
  \bibfield  {author} {\bibinfo {author} {\bibfnamefont {L.~C.~G.}\
  \bibnamefont {Govia}}, \bibinfo {author} {\bibfnamefont {G.~J.}\ \bibnamefont
  {Ribeill}}, \bibinfo {author} {\bibfnamefont {D.}~\bibnamefont {Rist{\`e}}},
  \bibinfo {author} {\bibfnamefont {M.}~\bibnamefont {Ware}},\ and\ \bibinfo
  {author} {\bibfnamefont {H.}~\bibnamefont {Krovi}},\ }\bibfield  {title}
  {\bibinfo {title} {Bootstrapping quantum process tomography via a
  perturbative ansatz},\ }\href {https://doi.org/10.1038/s41467-020-14873-1}
  {\bibfield  {journal} {\bibinfo  {journal} {Nature Communications}\ }\textbf
  {\bibinfo {volume} {11}},\ \bibinfo {pages} {1084} (\bibinfo {year}
  {2020})}\BibitemShut {NoStop}%
\bibitem [{\citenamefont {Kern}\ and\ \citenamefont {Soc}(1990)}]{Kern1990}%
  \BibitemOpen
  \bibfield  {author} {\bibinfo {author} {\bibfnamefont {W.}~\bibnamefont
  {Kern}}\ and\ \bibinfo {author} {\bibfnamefont {J.~E.}\ \bibnamefont {Soc}},\
  }\bibfield  {title} {\bibinfo {title} {{The Evolution of Silicon Wafer
  Cleaning Technology}},\ }\href {https://doi.org/10.1149/1.2086825} {\bibfield
   {journal} {\bibinfo  {journal} {J. Electrochem. Soc.}\ }\textbf {\bibinfo
  {volume} {137}},\ \bibinfo {pages} {1887} (\bibinfo {year}
  {1990})}\BibitemShut {NoStop}%
\bibitem [{\citenamefont {Dolan}(1977)}]{Dolan1977a}%
  \BibitemOpen
  \bibfield  {author} {\bibinfo {author} {\bibfnamefont {G.~J.}\ \bibnamefont
  {Dolan}},\ }\bibfield  {title} {\bibinfo {title} {{Offset masks for lift-off
  photoprocessing}},\ }\href {https://doi.org/10.1063/1.89690} {\bibfield
  {journal} {\bibinfo  {journal} {Appl. Phys. Lett.}\ }\textbf {\bibinfo
  {volume} {31}},\ \bibinfo {pages} {337} (\bibinfo {year} {1977})}\BibitemShut
  {NoStop}%
\bibitem [{\citenamefont {Chow}\ \emph {et~al.}(2014)\citenamefont {Chow},
  \citenamefont {Gambetta}, \citenamefont {Magesan}, \citenamefont {Abraham},
  \citenamefont {Cross}, \citenamefont {Johnson}, \citenamefont {Masluk},
  \citenamefont {Ryan}, \citenamefont {Smolin}, \citenamefont {Srinivasan},\
  and\ \citenamefont {Steffen}}]{Chow2014}%
  \BibitemOpen
  \bibfield  {author} {\bibinfo {author} {\bibfnamefont {J.~M.}\ \bibnamefont
  {Chow}}, \bibinfo {author} {\bibfnamefont {J.~M.}\ \bibnamefont {Gambetta}},
  \bibinfo {author} {\bibfnamefont {E.}~\bibnamefont {Magesan}}, \bibinfo
  {author} {\bibfnamefont {D.~W.}\ \bibnamefont {Abraham}}, \bibinfo {author}
  {\bibfnamefont {A.~W.}\ \bibnamefont {Cross}}, \bibinfo {author}
  {\bibfnamefont {B.~R.}\ \bibnamefont {Johnson}}, \bibinfo {author}
  {\bibfnamefont {N.~A.}\ \bibnamefont {Masluk}}, \bibinfo {author}
  {\bibfnamefont {C.~A.}\ \bibnamefont {Ryan}}, \bibinfo {author}
  {\bibfnamefont {J.~A.}\ \bibnamefont {Smolin}}, \bibinfo {author}
  {\bibfnamefont {S.~J.}\ \bibnamefont {Srinivasan}},\ and\ \bibinfo {author}
  {\bibfnamefont {M.}~\bibnamefont {Steffen}},\ }\bibfield  {title} {\bibinfo
  {title} {{Implementing a strand of a scalable fault-tolerant quantum
  computing fabric}},\ }\href {https://doi.org/10.1038/ncomms5015} {\bibfield
  {journal} {\bibinfo  {journal} {Nat. Commun.}\ }\textbf {\bibinfo {volume}
  {5}},\ \bibinfo {pages} {1} (\bibinfo {year} {2014})},\ \Eprint
  {https://arxiv.org/abs/1311.6330} {arXiv:1311.6330} \BibitemShut {NoStop}%
\bibitem [{\citenamefont {Ryan}\ \emph {et~al.}(2009)\citenamefont {Ryan},
  \citenamefont {Laforest},\ and\ \citenamefont
  {Laflamme}}]{ryan2009randomized}%
  \BibitemOpen
  \bibfield  {author} {\bibinfo {author} {\bibfnamefont {C.}~\bibnamefont
  {Ryan}}, \bibinfo {author} {\bibfnamefont {M.}~\bibnamefont {Laforest}},\
  and\ \bibinfo {author} {\bibfnamefont {R.}~\bibnamefont {Laflamme}},\
  }\bibfield  {title} {\bibinfo {title} {Randomized benchmarking of single-and
  multi-qubit control in liquid-state nmr quantum information processing},\
  }\href
  {https://iopscience.iop.org/article/10.1088/1367-2630/11/1/013034/meta}
  {\bibfield  {journal} {\bibinfo  {journal} {New J. Phys.}\ }\textbf {\bibinfo
  {volume} {11}},\ \bibinfo {pages} {013034} (\bibinfo {year}
  {2009})}\BibitemShut {NoStop}%
\bibitem [{\citenamefont {Ding}\ \emph {et~al.}(2019)\citenamefont {Ding},
  \citenamefont {Cui}, \citenamefont {Huang}, \citenamefont {Li}, \citenamefont
  {Tu},\ and\ \citenamefont {Guo}}]{Ding2019}%
  \BibitemOpen
  \bibfield  {author} {\bibinfo {author} {\bibfnamefont {Z.~H.}\ \bibnamefont
  {Ding}}, \bibinfo {author} {\bibfnamefont {J.~M.}\ \bibnamefont {Cui}},
  \bibinfo {author} {\bibfnamefont {Y.~F.}\ \bibnamefont {Huang}}, \bibinfo
  {author} {\bibfnamefont {C.~F.}\ \bibnamefont {Li}}, \bibinfo {author}
  {\bibfnamefont {T.}~\bibnamefont {Tu}},\ and\ \bibinfo {author}
  {\bibfnamefont {G.~C.}\ \bibnamefont {Guo}},\ }\bibfield  {title} {\bibinfo
  {title} {{Fast High-Fidelity Readout of a Single Trapped-Ion Qubit via
  Machine-Learning Methods}},\ }\href
  {https://doi.org/10.1103/PhysRevApplied.12.014038} {\bibfield  {journal}
  {\bibinfo  {journal} {Phys. Rev. Appl.}\ }\textbf {\bibinfo {volume} {12}},\
  \bibinfo {pages} {014038} (\bibinfo {year} {2019})},\ \Eprint
  {https://arxiv.org/abs/1810.07997} {arXiv:1810.07997} \BibitemShut {NoStop}%
\bibitem [{\citenamefont {Gely}\ \emph {et~al.}(2018)\citenamefont {Gely},
  \citenamefont {Steele},\ and\ \citenamefont {Bothner}}]{Gely2018}%
  \BibitemOpen
  \bibfield  {author} {\bibinfo {author} {\bibfnamefont {M.~F.}\ \bibnamefont
  {Gely}}, \bibinfo {author} {\bibfnamefont {G.~A.}\ \bibnamefont {Steele}},\
  and\ \bibinfo {author} {\bibfnamefont {D.}~\bibnamefont {Bothner}},\
  }\bibfield  {title} {\bibinfo {title} {Nature of the lamb shift in weakly
  anharmonic atoms: From normal-mode splitting to quantum fluctuations},\
  }\href {https://doi.org/10.1103/PhysRevA.98.053808} {\bibfield  {journal}
  {\bibinfo  {journal} {Phys. Rev. A}\ }\textbf {\bibinfo {volume} {98}},\
  \bibinfo {pages} {053808} (\bibinfo {year} {2018})}\BibitemShut {NoStop}%
\bibitem [{\citenamefont {Johnson}\ and\ \citenamefont {et~al.}(2021)}]{qgl}%
  \BibitemOpen
  \bibfield  {author} {\bibinfo {author} {\bibfnamefont {B.}~\bibnamefont
  {Johnson}}\ and\ \bibinfo {author} {\bibnamefont {et~al.}},\ }\href@noop {}
  {\bibinfo {title} {{Quantum Gate Language}}},\ \bibinfo {howpublished}
  {\url{https://github.com/BBN-Q/QGL}} (\bibinfo {year} {2021})\BibitemShut
  {NoStop}%
\bibitem [{\citenamefont {Ryan}\ \emph {et~al.}(2017)\citenamefont {Ryan},
  \citenamefont {Johnson}, \citenamefont {Ristè}, \citenamefont {Donovan},\
  and\ \citenamefont {Ohki}}]{Ryan2017}%
  \BibitemOpen
  \bibfield  {author} {\bibinfo {author} {\bibfnamefont {C.~A.}\ \bibnamefont
  {Ryan}}, \bibinfo {author} {\bibfnamefont {B.~R.}\ \bibnamefont {Johnson}},
  \bibinfo {author} {\bibfnamefont {D.}~\bibnamefont {Ristè}}, \bibinfo
  {author} {\bibfnamefont {B.}~\bibnamefont {Donovan}},\ and\ \bibinfo {author}
  {\bibfnamefont {T.~A.}\ \bibnamefont {Ohki}},\ }\bibfield  {title} {\bibinfo
  {title} {Hardware for dynamic quantum computing},\ }\href
  {https://doi.org/10.1063/1.5006525} {\bibfield  {journal} {\bibinfo
  {journal} {Review of Scientific Instruments}\ }\textbf {\bibinfo {volume}
  {88}},\ \bibinfo {pages} {104703} (\bibinfo {year} {2017})}\BibitemShut
  {NoStop}%
\bibitem [{\citenamefont {Ryan}\ and\ \citenamefont {et~al.}(2016)}]{QDSP}%
  \BibitemOpen
  \bibfield  {author} {\bibinfo {author} {\bibfnamefont {C.}~\bibnamefont
  {Ryan}}\ and\ \bibinfo {author} {\bibnamefont {et~al.}},\ }\href@noop {}
  {\bibinfo {title} {{BBN-QDSP-X6}}},\ \bibinfo {howpublished}
  {\url{https://github.com/BBN-Q/BBN-QDSP-X6}} (\bibinfo {year}
  {2016})\BibitemShut {NoStop}%
\bibitem [{\citenamefont {Rowlands}\ and\ \citenamefont
  {et~al.}(2021)}]{auspex}%
  \BibitemOpen
  \bibfield  {author} {\bibinfo {author} {\bibfnamefont {G.}~\bibnamefont
  {Rowlands}}\ and\ \bibinfo {author} {\bibnamefont {et~al.}},\ }\href@noop {}
  {\bibinfo {title} {{Auspex}}},\ \bibinfo {howpublished}
  {\url{https://github.com/BBN-Q/Auspex}} (\bibinfo {year} {2021})\BibitemShut
  {NoStop}%
\bibitem [{\citenamefont {Blume-Kohout}\ \emph {et~al.}(2021)\citenamefont
  {Blume-Kohout}, \citenamefont {da~Silva}, \citenamefont {Nielsen},
  \citenamefont {Proctor}, \citenamefont {Rudinger}, \citenamefont {Sarovar},\
  and\ \citenamefont {Young}}]{blumekohout2021taxonomy}%
  \BibitemOpen
  \bibfield  {author} {\bibinfo {author} {\bibfnamefont {R.}~\bibnamefont
  {Blume-Kohout}}, \bibinfo {author} {\bibfnamefont {M.~P.}\ \bibnamefont
  {da~Silva}}, \bibinfo {author} {\bibfnamefont {E.}~\bibnamefont {Nielsen}},
  \bibinfo {author} {\bibfnamefont {T.}~\bibnamefont {Proctor}}, \bibinfo
  {author} {\bibfnamefont {K.}~\bibnamefont {Rudinger}}, \bibinfo {author}
  {\bibfnamefont {M.}~\bibnamefont {Sarovar}},\ and\ \bibinfo {author}
  {\bibfnamefont {K.}~\bibnamefont {Young}},\ }\href@noop {} {\bibinfo {title}
  {A taxonomy of small markovian errors}} (\bibinfo {year} {2021}),\ \Eprint
  {https://arxiv.org/abs/2103.01928} {arXiv:2103.01928 [quant-ph]} \BibitemShut
  {NoStop}%
\end{thebibliography}%


\FloatBarrier % so figures aren't in the bibliography
\newpage
\onecolumngrid
\newpage
\section{SUPPLEMENTAL MATERIAL} 
\section{Device Parameters}\label{app:device}

The superconducting transmon device was fabricated by BBN in collaboration with Raytheon RF Components. The device ground plane, resonators and qubit capacitors are \SI{200}{\nm} niobium sputtered on high-resistivity intrinsic silicon, cleaned with an HF-last RCA clean \cite{Kern1990} before sputtering. The niobium metallization was optically patterned and etched with an \ch{SF6}+\ch{O2} RIE-ICP plasma etch. Post-etch residues were removed using an oxygen ash and a \ch{HF} etch. The qubits' single Josephson junction was patterned using a Dolan bridge \cite{Dolan1977a} technique using a \ch{PMMA}-\ch{MMA} bilayer resist and electron beam lithography. The junction was fabricated using aluminum electron beam evaporation after an \ch{Ar+} ion mill etch to remove surface oxides. The sample was mounted in and wirebonded to a custom copper sample holder, with additional aluminum wirebonds across on-chip resonators to short parasitic resonances. This package was in turn mounted to the cold stage of a dilution refrigerator inside a light-tight, magnetically shielded sample can.

\begin{figure}
  \includegraphics[width=.6\textwidth]{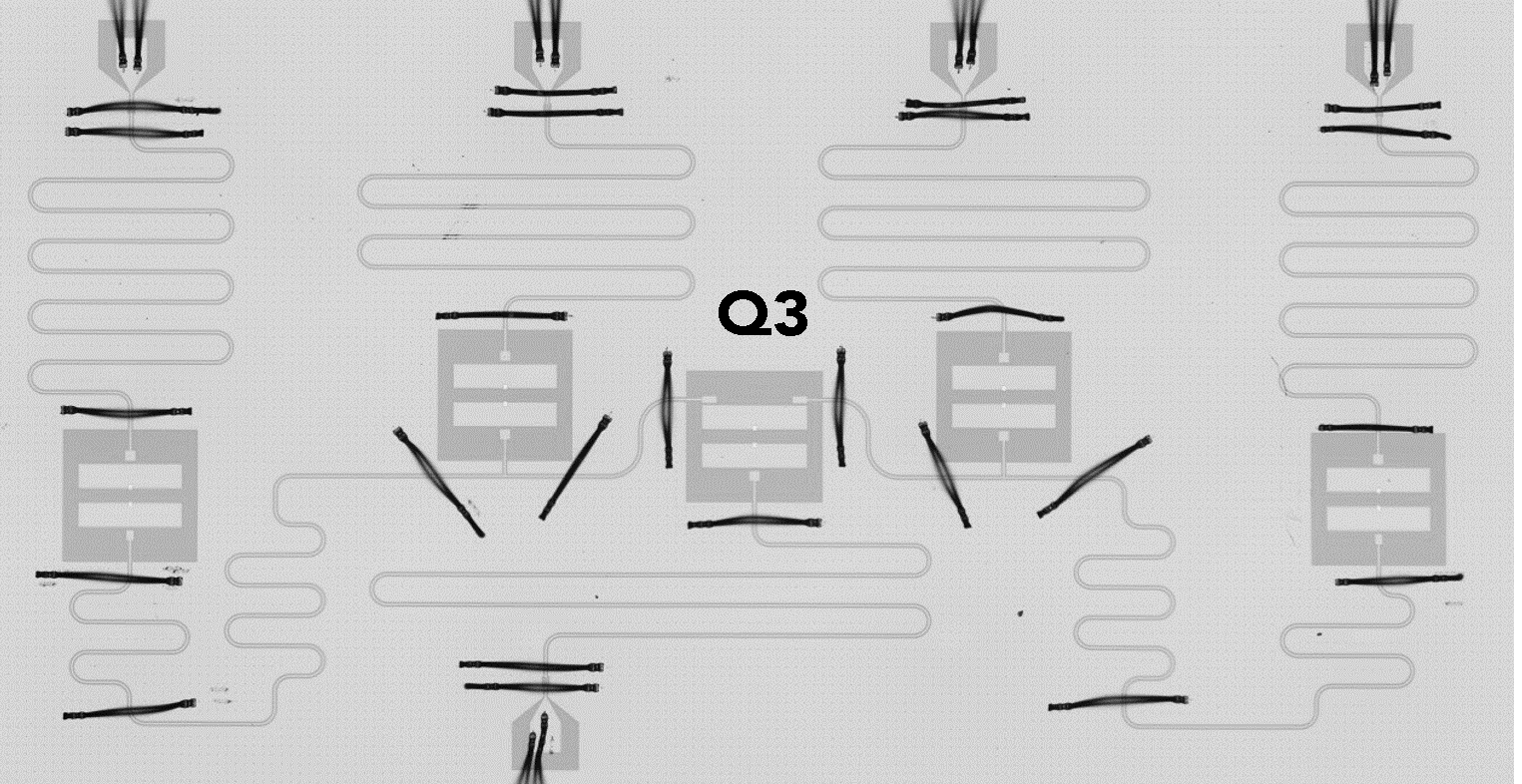}
  \caption{Micrograph showing the device studied in this Letter. Q\textsubscript{3} is the central qubit.\label{fig:device}}
\end{figure}

\begin{figure}
  \includegraphics[width=.8\textwidth]{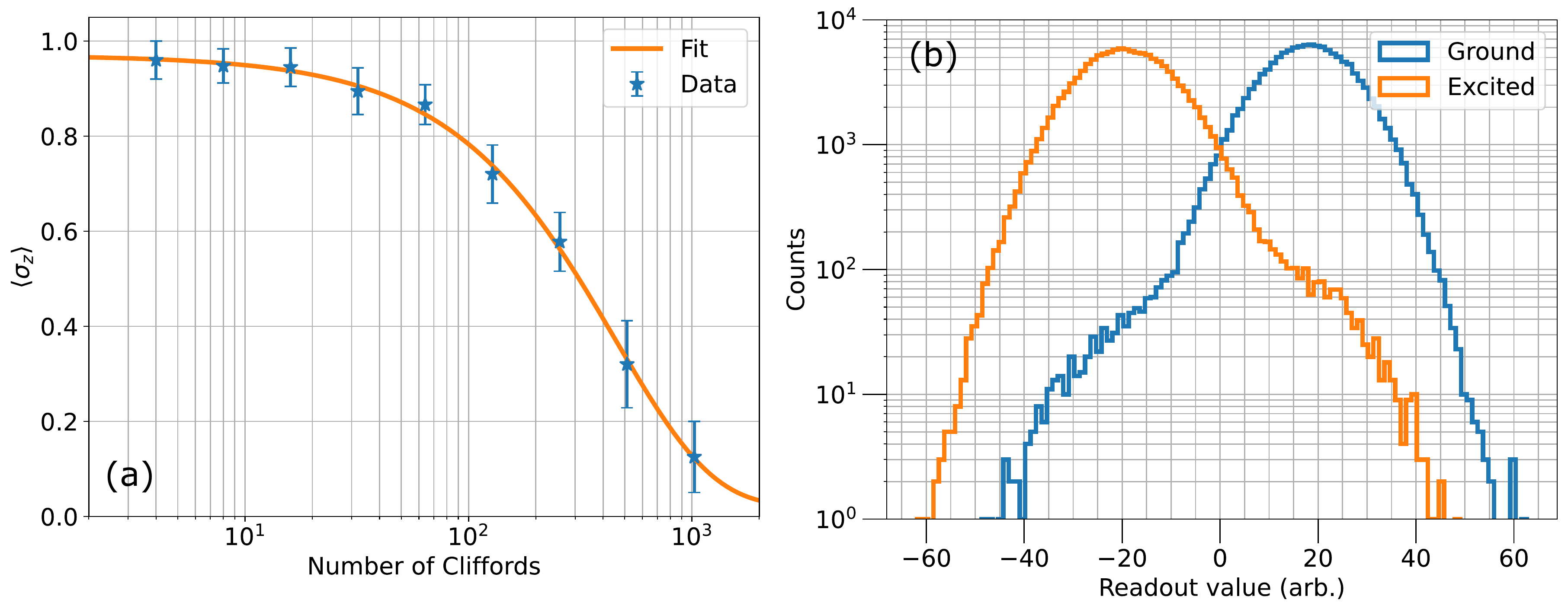}
  \caption{(a) Randomized benchmarking of single-qubit Clifford gates on Q\textsubscript{3}. Cliffords are generated from $\{I, X(\pm\pi/2), Y(\pm\pi/2), X(\pi), Y(\pi)\}$ gates with an average of \num{1.71} gates per Clifford. Points are averages of 32 independent randomized sequences of Clifford gates for each length, while the solid curve is an exponential fit to the data used to extract the error per Clifford $r = \num{1.1e-3}$. (b) Histogram of measurement results after matched filter integration for \num{1.3e5} ground and excited state preparations, corresponding to $F = \SI{96.4}{\percent}.$  \label{fig:qb_char}}
\end{figure}

The qubit chip consists of five fixed-frequency transmon qubits, designed to be similar to those described in \cite{Chow2014}, connected by bus resonators in two pairs of three. A micrograph of the device is shown in Figure \ref{fig:device}. For the experiments described in this Letter, only one qubit (Q\textsubscript{3}) is measured, while the other transmons are detuned by at least $\SI{140}{\mega\hertz}$ (with coupling only through bus resonators) and so have no impact on its operation and can thus be safely ignored. Q\textsubscript{3} is dispersively coupled to a readout resonator through which control drives resonant with the qubit are also applied. A detailed description of the control wiring, electronics and software stack can be found in the~\hyperref[app:electronics]{Control Electronics} section. Relevant device parameters are listed in Table \ref{tab:qbparams}. In particular, the photon number population evolution in the qubit cavity and its relaxation time $1/\kappa$ were measured using the Stark shift \cite{McClure2016}. Qubit coherences were measured using standard inversion recovery, Ramsey and Hahn echo sequences, and are listed in Table \ref{tab:coherence}. The $X_{\tfrac{\pi}{2}}$ and $Y_{\tfrac{\pi}{2}}$ qubit rotation gates were implemented as Gaussian pulses with a \SI{60}{\ns} length. Single-qubit error per Clifford gate was measured using randomized benchmarking \cite{ryan2009randomized} and found to be $r = \num{1.1e-3}$ (Fig. \ref{fig:qb_char}a), consistent with the results of GST (Fig. \ref{fig:diamond}). Qubit measurement fidelity, here defined as $F = (P_{0|0} + P_{1|1})/2$, where $P_{1|1}$ is the probability of correctly identifying the qubit state as $\ket{1}$ when prepared in $\ket{1}$, was determined from calibration data taken simultaneously with the QILGST sequences.  To calibrate the measurement fidelity, we used \num{1.3e5} preparations each of the qubit in its ground and excited states. The reflected cavity signal was downconverted and integrated using a matched kernel filter \cite{ryan2015tomography}, and binned resulting in the well-separated readout histograms shown in Figure \ref{fig:qb_char}b. Integrating and taking the difference of these histograms yields a fidelity $F = \SI{96.35}{\percent}$, while an approach using logistic regression \cite{Ding2019} yields a fidelity $F = \SI{96.43}{\percent} \pm \SI{0.7}{\percent} $.

\begin{table}
\centering
\begin{tabular}{|c|c|c|c|}
    \hline
     Parameter & Symbol & Value & Measurement  \\ \hhline{|=|=|=|=|}
     Qubit frequency & $\omega_{01}/2\pi$ & \SI{4.76418}{\giga\hertz} & Low-power qubit spectroscopy \\ \hline
     Qubit anharmonicity & $\alpha$ & \SI{310}{\mega\hertz} & Two-tone qubit spectroscopy \\ \hline
     Resonator dressed frequency & $\omega_r/2\pi$ & \SI{6.73464}{\giga\hertz} & Low power resonator spectroscopy \\ \hline
     Resonator-qubit coupling & $g/2\pi$ & \SI{53.4}{\mega\hertz} & Calculated \cite{Gely2018} \\ \hline
     Qubit dispersive shift & $\chi/2\pi$ & \SI{-0.270}{\mega\hertz} & Resonator spectroscopy with qubit in $\ket{0}$ and $\ket{1}$ \\ \hline
     Resonator photon decay rate & $1/\kappa$ & \SI{242}{\ns} & Cavity photon number decay \cite{McClure2016} \\ \hline
\end{tabular}
\caption{Device parameters for transmon Q\textsubscript{3}. \label{tab:qbparams}}
\end{table}

\begin{table}
    \centering
    \begin{tabular}{|c|c|c|}
         \hline
         $T_1$ (\si{\micro\second}) & $T_2^*$ (\si{\micro\second}) & $T_2$ (\si{\micro\second})  \\ \hhline{|=|=|=}
         70.2 & 43.8 & 82.5 \\ \hline
    \end{tabular}
    \caption{Transmon average coherence times, measured continuously over \SI{8}{\hour}.}
    \label{tab:coherence}
\end{table}

\section{Control electronics}\label{app:electronics}

QILGST control sequences are generated in the \texttt{pyGSTi} software package then compiled and time-ordered using BBN's Quantum Gate Language (QGL)~\cite{qgl}. QGL ouputs a hardware efficient representation of the experiments which are sent to the control hardware over an ethernet interface. The physical control and readout pulses are sequenced using BBN's custom Arbitrary Pulse Sequencer II (APS-II). The sequencing capabilities of the APS-II allow for continuous playback of the QILGST experiments in a interleaved fashion collecting 1024 shots for each $t_d$ without interruption for waveform or data loading.

\begin{figure}
  \includegraphics[width=.8\textwidth]{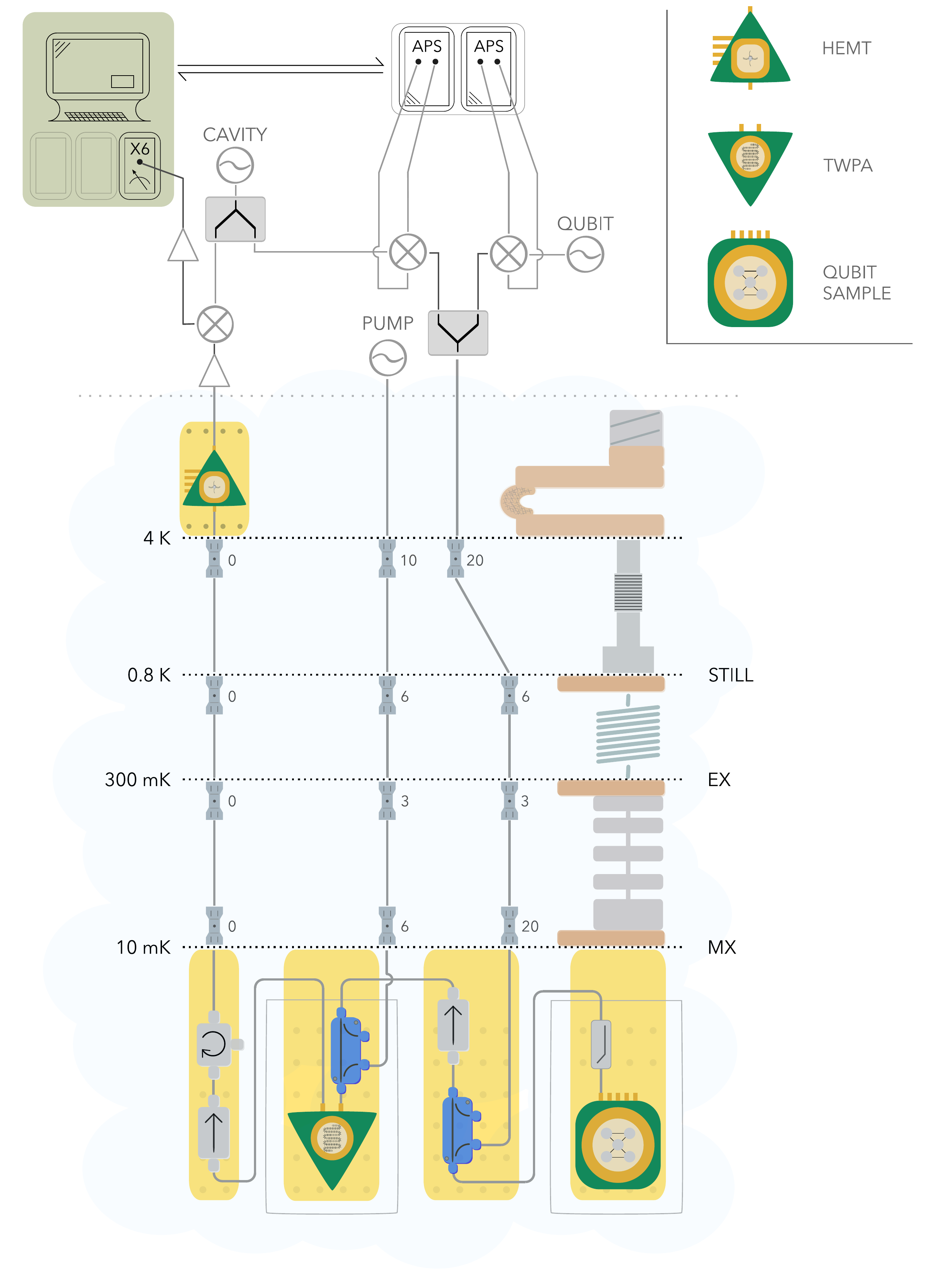}
  \caption{Experimental diagram. Microwave control signals are synthesized at room temperature and mixed up to the qubit and cavity frequencies. These signals are routed into a dilution refrigerator and bounce off the qubit sample. The microwaves then pass through both a JTWPA and HEMT amplifier before being down-mixed and further amplified again at room temperature. Microwave attenuation levels are listed for each temperature stage of the dilution refrigerator. The measurement signal from the cavity is converted to an intermediate frequency using an autodyne technique~\cite{ryan2015tomography} and digitized using an commercially available digitzer running custom firmware~\cite{QDSP}.  \label{fig:circuit_diagram}}
\end{figure}

Our superconducting device is measured in a Bluefors LD-40 dilution refrigerator. Fig.~\ref{fig:circuit_diagram} outlines the complete measurement system. The amplifier pump and qubit control and readout microwave tones are generated using Holzworth9000A microwave synthesizers. To correct for any residual phase instability in the measurement tone, we use an \lq autodyne\rq measurement technique~\cite{ryan2015tomography}. Control and readout pulses are mixed with the microwave tones using Marki IQ-4509 mixers. Control pulses are generated by BBN custom Arbitrary Pulse Sequencer-II (APS-II)~\cite{Ryan2017} units. The readout and control channels are combined at room temperature, and the qubit cavity is measured in reflection through a Krytar directional coupler at the cold stage. A K\&L micro machined 6L250 low-pass filter provides the qubit with protection from high frequency noise above 12 GHz, and a Quinstar QCI cryogenic isolator provides further isolation from the rest of the readout chain. The cavity signal and a pump tone are then combined using a second directional coupler and sent through a Josephson Traveling-Wave Parametric Amplifier (JTWPA). The JTWPA provides roughly 25 dB of gain at the cavity frequency. Additional isolation is provided by a second Krytar QCI isolator and QCY circulator. The readout signal is then amplified at the 4 K stage using an LNF LNC4\_8C HEMT amplifier.

Outside the cryostat, microwaves are amplified further using a L3Harris Narda-MITEQ AMF-4F-04001200-15-10P before downconversion to the \SI{13}{\mega\hertz} intermediate frequency with a Marki doubly balanced mixer. A Stanford Research Systems SR445A preamplifier with a voltage gain of 25 is the last stage of amplification before the signal is captured using a X6-1000M Innovative Integration digitizer card running custom firmware~\cite{QDSP} which further decimates, digitally downconverts and integrates the data using a matched filter \cite{ryan2015tomography}. Data collection and pipelining is orchestrated by a the Auspex software package~\cite{auspex}. All sources, digitizers and sequencers share a global 10 MHz clock provided by an SRS SF725 Rubidium frequency standard. 

\section{QILGST Simulations}
In the main text, we demonstrated that QILGST worked correctly using simulated data. Here we provide the details of these simulations. We simulated single-qubit QILGST on a gate set consisting of $\nicefrac{\pi}{2}$ rotations around the $\sigmax$ and $\sigmay$ axes, mid-circuit and terminating measurements in the $\sigmaz$ basis, and state preparation in $\ket{0}$. The target QI is $\gate{Q}_{\text{target}} =\{\gate{Q}_{\text{target},0},\gate{Q}_{\text{target},1}\}$ where
\begin{equation}
\gate{Q}_{\text{target},k}[\rho] =  \text{Tr}\left[\frac{1}{2}\left(\sigmai + (-1)^k\sigmaz\right)\rho\right]\left(\sigmai + (-1)^k\sigmaz\right).
\label{eq:zQI1}
\end{equation}
We generated 100 different error models. Errors on the mid-circuit measurement were randomly sampled, and errors on all other operations were held constant. The X and Y operations each had over-rotations of $10^{-3}$ radians along both X and Y axes and were subject to $10^{-2}$ depolarization, while SPAM was subject to $10^{-3}$ depolarization.  The mid-circuit measurement was subject to randomly chosen errors, both on the classical and quantum portions of the channel.  With a probability chosen uniformly from 0 to $10^{-2}$ the $|0\rangle$ was misidentified as $|1\rangle$ (and vice-versa, with another independently chosen probability).  Additionally, X and Y coherences were chosen to persist post-mid-circuit measurement, both strengths equal but chosen uniformly at random between $0$ and $10^{-2}$.

For each error model, we simulated drawing $N$ samples from each of the QILGST circuits, with $N$ varying logarithmically from 16 to 1024. For the data with each value of $N$, we applied the QILGST analysis to obtain an estimate of the gate set. We then computed half the diamond distance between the estimated QI $\widehat{\gate{Q}}$ and the \emph{true} QI $\gate{Q}_{\text{true}}$ used in the simulation (\emph{not} $\gate{Q}_{\text{target}}$), as a measure of the estimation inaccuracy. Fig.~\ref{fig:simulations} shows estimation inaccuracy versus $N$. It scales as $\nicefrac{1}{\sqrt{N}}$ (standard quantum-limited scaling), indicating that QILGST is correctly reconstructing the QI up to the expected statistical fluctuations.

\section{QILGST Experimental Results}
\begin{figure}
\includegraphics[width=8 cm]{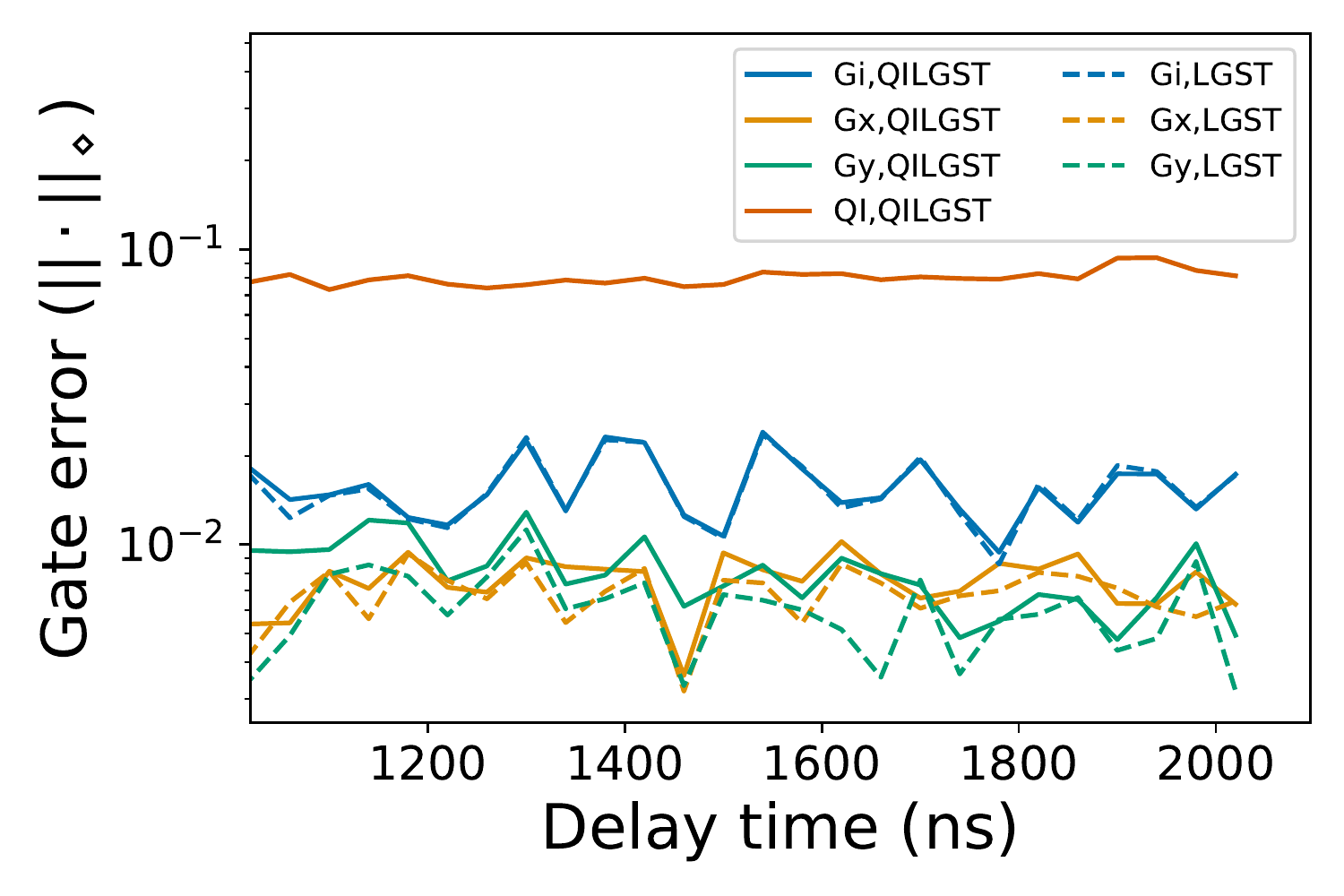}
\caption{\label{fig:diamond} Errors as measured by half diamond distance for QILGST (solid lines) and LGST (dashed lines) reconstructions for $t_d>1020 ns$. $G_i$ denotes the idle operation, $G_x$ the $\pi/2$ X rotation, $G_y$ the $\pi/2$ Y rotation, and $Q$ the quantum instrument.}
\end{figure}

Here we include some additional analysis of the QILGST experimental results. Fig~\ref{fig:diamond} shows the the half diamond distance error ($\epsilon_{\diamond}$) for each gate for $t_d \geq 1020$ (when the datasets are Markovian). We examine the estimates obtained from both QILGST and LGST.  (The latter does not incorporate circuits containing any mid-circuit measurements, and therefore does not reconstruct an estimate for the quantum instrument.). There is good agreement between the LGST and QILGST reconstructions of the non-QI operations, and the error rates are reasonably stable across the examined delay times.

We also provide the QILGST reconstruction of $\widehat{\mathcal{G}}(\SI{2020}{ns})$ in Table \ref{tab:gatevals}.

\begin{table}
\centering
\begin{tabular}{|c|c|c|c|}
    \hline
     Operation label & $\mathcal{G}_\text{target}$ & $\widehat{\mathcal{G}}(\SI{2020}{ns})$ & $2\sigma$ error bars  \\ \hhline{|=|=|=|=|}
     $|\rho\rangle\rangle$ & $\frac{1}{\sqrt{2}} \left(\begin{array}{c} 1\\0\\0\\1\end{array}\right)$ & $\left(\begin{array}{c} 1\\-0.016\\ -0.008\\0.953\end{array}\right)$ & $\left(\begin{array}{c} 0\\0.008\\ 0.009\\0.006\end{array}\right)$ \\ \hline
     $\langle\langle M|$ & $\frac{1}{\sqrt{2}} \left(\begin{array}{c c c c} 1 & 0 & 0 & 1\end{array}\right)$ & $\frac{1}{\sqrt{2}}\left(\begin{array}{c c c c} 1.002 & -0.002 & -0.01 & 0.997\end{array}\right)$ & $\left(\begin{array}{c c c c} 0.002 & 0.006 & 0.008 & 0.003\end{array}\right)$ \\ \hline
     $G_i$ & $\left(\begin{array}{c c c c}
1 & 0 & 0 & 0
\\
0 & 1 & 0 & 0
\\
0 & 0 & 1 & 0
\\
0 & 0 & 0 & 1
\end{array}\right)$ & 
$\left(\begin{array}{c c c c}
1.0 & 0.0 & 0.0 & 0.0
\\
-0.004 & 0.993 & -0.001 & 0.021
\\
0.01 & 0.008 & 0.989 & -0.008
\\
0.005 & -0.022 & 0.003 & 0.99
\end{array}\right)$
& 
$\left(\begin{array}{c c c c}
0.0 & 0.0 & 0.0 & 0.0
\\
0.008 & 0.009 & 0.024 & 0.027
\\
0.008 & 0.024 & 0.009 & 0.03
\\
0.009 & 0.027 & 0.03 & 0.01
\end{array}\right)$
\\ \hline
$G_x$ & 
$\left(\begin{array}{c c c c}
1 & 0 & 0 & 0
\\
0 & 1 & 0 & 0
\\
0 & 0 & 0 & -1
\\
0 & 0 & 1 & 0
\\
\end{array}\right)$
&
$\left(\begin{array}{c c c c}
1.0 & 0.0 & 0.0 & 0.0
\\
-0.001 & 0.999 & 0.003 & -0.004
\\
0.0 & -0.004 & 0.011 & -0.999
\\
0.0 & -0.003 & 0.999 & 0.011
\\
\end{array}\right)$
&
$\left(\begin{array}{c c c c}
0.0 & 0.0 & 0.0 & 0.0
\\
0.004 & 0.004 & 0.012 & 0.012
\\
0.002 & 0.012 & 0.007 & 0.003
\\
0.002 & 0.012 & 0.003 & 0.006
\\
\end{array}\right)$\\\hline
$G_y$ & 
$\left(\begin{array}{c c c c}
1 & 0 & 0 & 0
\\
0 & 0 & 0 & 1
\\
0 & 0 & 1 & 0
\\
0 & -1 & 0 & 0
\\
\end{array}\right)$
&
$\left(\begin{array}{c c c c}
1.0 & 0.0 & 0.0 & 0.0
\\
-0.001 & 0.005 & 0.004 & 0.999
\\
-0.001 & -0.005 & 0.999 & -0.004
\\
0.001 & -0.999 & -0.005 & 0.006
\\
\end{array}\right)$
&
$\left(\begin{array}{c c c c}
0.0 & 0.0 & 0.0 & 0.0
\\
0.003 & 0.006 & 0.013 & 0.003
\\
0.004 & 0.013 & 0.003 & 0.013
\\
0.003 & 0.003 & 0.013 & 0.006
\\
\end{array}\right)$
\\
\hline
$Q_0$ & $\left(\begin{array}{c c c c}
0.5 & 0 & 0 & 0.5
\\
0 & 0 & 0 & 0
\\
0 & 0 & 0 & 0
\\
0.5 & 0 & 0 & 0.5
\\
\end{array}\right)$
&
$\left(\begin{array}{c c c c}
0.504 & 0.003 & -0.006 & 0.493
\\
-0.01 & 0.002 & 0.005 & -0.014
\\
-0.007 & -0.005 & 0.002 & -0.0
\\
0.454 & 0.0 & 0.005 & 0.478
\\
\end{array}\right)$
&
$\left(\begin{array}{c c c c}
0.003 & 0.011 & 0.011 & 0.005
\\
0.013 & 0.023 & 0.023 & 0.016
\\
0.013 & 0.023 & 0.022 & 0.016
\\
0.006 & 0.014 & 0.014 & 0.009
\\
\end{array}\right)$
\\
\hline
$Q_1$  & $\left(\begin{array}{c c c c}
0.5 & 0 & 0 & -0.5
\\
0 & 0 & 0 & 0
\\
0 & 0 & 0 & 0
\\
-0.5 & 0 & 0 & 0.5
\\
\end{array}\right)$
&
$\left(\begin{array}{c c c c}
0.496 & -0.003 & 0.006 & -0.493
\\
0.004 & 0.001 & 0.001 & -0.009
\\
0.009 & -0.003 & -0.005 & -0.009
\\
-0.418 & 0.004 & 0.0 & 0.448
\\
\end{array}\right)$
&
$\left(\begin{array}{c c c c}
0.003 & 0.011 & 0.011 & 0.005
\\
0.013 & 0.023 & 0.023 & 0.015
\\
0.013 & 0.023 & 0.023 & 0.015
\\
0.007 & 0.015 & 0.016 & 0.01
\\
\end{array}\right)$ \\ \hline
\end{tabular}
\caption{QILGST-reconstructed estimates for all operations at $t_d=\SI{2020}{ns}$ . \label{tab:gatevals}}
\end{table}

\section{Stark Shift Model for Non-Markovian Error}

The QILGST fits in the main text show evidence for a considerable amount of non-Markovian error. A possible model to explain this error is the AC-Stark shift of the qubit frequency due to residual photons in the measurement cavity leftover from the mid-circuit measurement. The AC-Stark shift will be time-dependent as the photons leak out of the cavity, and thus induce a non-Markovian error on subsequent gates following the mid-circuit measurement. A qubit-only Hamiltonian describing this error model is given by
\begin{align}
    H(t) = H_{k} + \delta(t) \sigma_z,
\end{align}
where $H_{k}$ for $k \in \{x,y\}$ is the Hamiltonian describing the intended gate, and $\delta$ is a parameter describing the Stark-shift. Based on the lowest order dispersive theory, we would expect that $\delta(t) = \chi n(t)$, where $n(t) = \left<\hat{a}^\dagger\hat{a}\right>(t) = n_{i}e^{-\kappa t}$ is the time-dependent expectation value of the cavity photon population, with $n_{i}$ the initial photon population that depends on the outcome of the mid-circuit measurement, labeled by $i \in \{0,1\}$, as we drive on one of the shifted cavity lines for measurement. All parameters in this model have been measured by independent calibration experiments, see Table \ref{tab:qbparams}.

The gate generated by this Hamiltonian is given by
\begin{align}
    G_{k}(t) = \mathcal{T}_{\leftarrow}\exp\left(-i\int_{t_0}^{t_0 + t_{\rm gate}} H(t) dt\right) \approx \exp\left(-i\int_{t_0}^{t_0 + t_{\rm gate}} H(t) dt\right) = \exp\left( -i\left[H_{k}t_{\rm gate} + \varphi_{i,m} \sigma_z\right]\right),
\end{align}
where the approximation sign is an indication that we have approximated the full time-ordered integral with the first order term of the Magnus expansion. We have verified that the second order term of the Magnus expansion results in a phase error that is at least an order of magnitude smaller than the first order phase error $\varphi_{i,m}$. From the experimental calibration and dispersive theory, the first order phase error is given by
\begin{align}
    \varphi_{i,m}(t_d) = \frac{\chi}{n_{i}}\left(1 - e^{-\kappa t_{\rm gate}}\right)e^{-\kappa\left(mt_{\rm gate} + t_{d}\right)}, \label{eqn:phiBBN}
\end{align}
where $m \in \{0,1,2\}$ labels the gates following the mid-circuit measurement.

For the modelling results presented in the main text, we approximate the implemented gate more accurately by replacing $H_{k}$ with $\log(\widehat{G}_k)$, the generator \cite{blumekohout2021taxonomy} of the superoperator representation of the gate $\widehat{G}_k$ characterized by QILGST at 2020ns delay. We model each gate following the mid-circuit measurement as
\begin{align}
\widehat{G}_{k}(\alpha, r, i, m) = \exp[\log(\widehat{G}_{k}) + \alpha_i(t_d) \exp(-mr_i) \mathcal{Z}], \label{eqn:StarkModel}
\end{align}
where
\begin{align}
    \alpha_i(t_d) = \frac{\chi}{n_{i}}\left(1 - e^{-\kappa t_{\rm gate}}\right)e^{-\kappa t_{d}},
\end{align}
and $r_i = \kappa t_{\rm gate}$ under the dispersive model. 

In addition to the model given above with $\varphi_{i,m}$ of Eq.~\eqref{eqn:phiBBN} determined entirely by independently characterized parameters, we also fit a model of the form of Eq.~\eqref{eqn:StarkModel} with all $\alpha_i(t_d)$ and $r_i$ as free parameters. For this model fitting, these free parameters are independently fit at each delay time, and for each mid-circuit measurement result. The model fit results in a phase error
\begin{align}
    \phi_{i,m}(t_d) = \alpha_i(t_d)e^{-mr_i}.
\end{align}

\begin{figure}[h!]
\includegraphics[width=15cm]{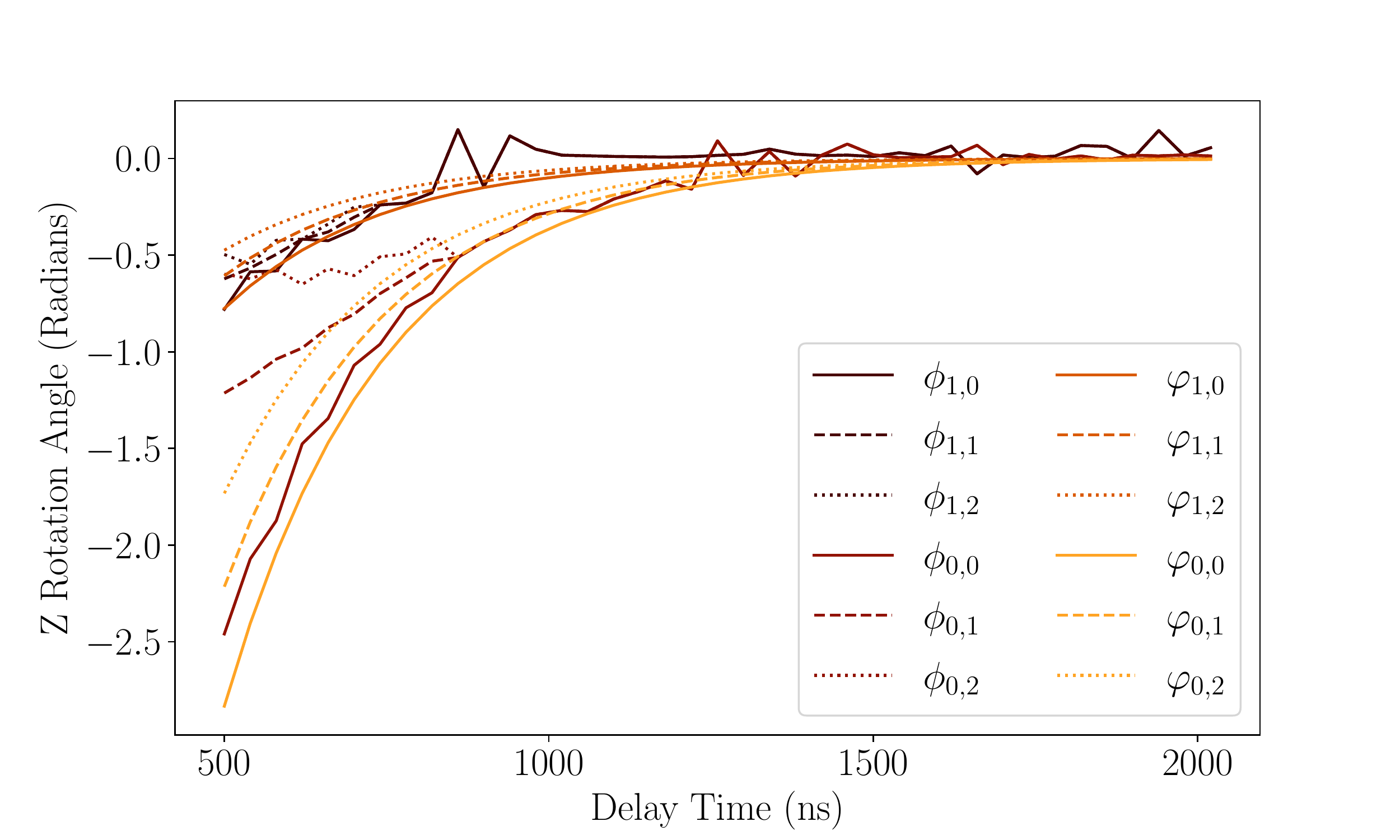}
\caption{The total amount of $\sigma_z$ rotation angle induced by the post-measurement Stark shift in each post-measurement gate.  $\varphi_{i,m}$ indicates the theoretical prediction for the $m^{th}$ post-measurement gate when $i$ is read out; $\phi_{i,m}$ is the same quantity, but where we numerically optimized the fit parameters of the model (independently at each $t_d$).}
\label{fig:z-rotation}
\end{figure}

Fig.~\ref{fig:z-rotation} shows the calculated value of $\varphi_{i,m}$, and the fit estimate of $\phi_{i,m}$ as a function of delay time for each gate following the mid-circuit measurement. The two models agree reasonably well above $t_d\approx900ns$.  We note that a) the effect is much more pronounced (as expected) when 0 is read out than 1, and b) the fit model does not quite follow an exponential decay at low $t_d$. This latter point indicates that while the Stark shift inspired model of Eq.~\eqref{eqn:StarkModel} with free fit parameters is a good effective model for the data, it does not agree with simple microscopic dispersive theory. This indicates that the discrepancy is likely not due to mis-characterization of the system parameters, but of qualitatively distinct physics arising from effects outside of dispersive theory.

\end{document}